\documentclass[sigconf]{acmart}
\AtBeginDocument{%
  \providecommand\BibTeX{{%
    \normalfont B\kern-0.5em{\scshape i\kern-0.25em b}\kern-0.8em\TeX}}}

\setcopyright{acmcopyright}
\copyrightyear{2021}
\acmYear{2021}

\acmConference[]{}

\usepackage[ruled,vlined]{algorithm2e}
\SetKwInput{KwInput}{Input}                
\SetKwInput{KwOutput}{Output}              
\SetKwFor{RepTimes}{repeat}{times}{until}

\SetAlgoLined
\SetKwProg{Fn}{Function}{ is}{end}

\SetKwProg{Def}{def}{:}{}

\usepackage{enumitem}
\usepackage{array}
\usepackage{soul}

\newcolumntype{L}[1]{>{\raggedright\let\newline\\\arraybackslash\hspace{0pt}}m{#1}}

\newcommand{\red}[1]{\textcolor{black}{#1}}
\newcommand{\blue}[1]{{#1}}

\newcommand{\hush}[1]{}
\usepackage{graphicx}

\renewcommand\footnotetextcopyrightpermission[1]{} 

\begin{document}

\title{Prioritizing Original News on Facebook}

\author{Xiuyan Ni}
\email{xni@fb.com}
\affiliation{%
  \institution{Facebook Inc.}
}

\author{Shujian Bu}
\email{shujian@fb.com}
\affiliation{%
  \institution{Facebook Inc.}
}

\author{Igor L. Markov}
\email{imarkov@fb.com}
\affiliation{%
  \institution{Facebook Inc.}
}

\renewcommand{\shortauthors}{}

\begin{abstract}
This work outlines how we prioritize \textit{original} news, a critical indicator of news quality. By examining the landscape and life-cycle of news posts on our social media platform, we identify challenges of building and deploying an originality score. We pursue an approach based on normalized PageRank values and three-step clustering, and refresh the score on an hourly basis to capture the dynamics of online news. We describe a near real-time system architecture, 
evaluate our methodology, and deploy it to production. Our empirical results validate individual components and show that
prioritizing original news increases user engagement with news and improves proprietary cumulative metrics. 

\end{abstract}



\maketitle

\section{Introduction}

Large amounts of news are published online every day, and many people now primarily consume news online \cite{reis2015breaking}. News quality affects how people consume news and which platforms they prefer \cite{facebook,news2021fb}. Expressing news quality numerically can facilitate significant improvements for users and platforms \cite{publishers2019fb}. Among various aspects of news quality, we focus on \textit{originality}, which can be contrasted with duplicates, slightly edited text, and coverage that references original news. Producing original news is laborious and requires expertise, but
such efforts initiate the typical news cycle and drive the entire news industry.
Original news inform people around the world, from breaking news, eye-witness reports and critical updates at the time of crisis, to in-depth investigative reports that uncover new facts. Prioritizing original news online is in everyone's long-term interest~\cite{publishers2019fb}.

\blue{In this work, we first explore the landscape of online news, using the Facebook platform as an example. To enable a quantitative approach, we tabulate the spectrum of news originality from \textit{completely unoriginal} to \textit{highly original} news. Our static analysis suggests that highly original news are rare, despite a large inventory which needs to be indexed and processed to accurately identify the original ones. We also explore the dynamics of the news life-cycle on Facebook and find that news posts typically attain the greatest exposure in the first couple of hours, followed by a long tail. This result suggests that an originality score used to improve News Feed ranking must be computed promptly.}

\blue{Given two challenges --- search quality at scale and fast response --- we build a near real-time system and construct a synthesized signal for news originality. News articles that cover the same news event are clustered together based on specialized BERT embeddings \cite{devlin2018bert}, which are finetuned on pairwise-labeled data (same subject or different subjects). After evaluating several clustering algorithms against human-labeled pairwise data, we settle on a two-stage clustering algorithm that is both effective and highly scalable to large datasets. To adequately capture news dynamics, our system performs incremental updates on an hourly basis.}

\blue{We concluded that content alone is insufficient to judge news originality, but behavioral signals such as citations of prior posts can also be used. Integrity considerations are particularly important, given the high  incentives to game online news distribution. To de-bias our algorithms, we filter out news articles produced within patterns of nefarious activity.
We first evaluate the performance of our originality signal offline against ratings by professional journalists. Online evaluation is based on an A/B test where we additionally monitor the impact on news article ranking~\cite{ab2015kdd}. The signal 
is incorporated in the News Feed ranking system.} \\

Our contributions include: 
\begin{itemize}
\item We examine the news originality landscape and the dynamics of the news life-cycle, then propose a quantitative approach to reason about the news ecosystem. We categorize the level of news originality by the effort spent to generate news content. 
\item We propose a methodology and architect a near real-time system that processes individual news articles at a large scale. Using the PageRank algorithm and three-step clustering, it calculates a synthetic score to estimate news originality. PageRank
normalization within clusters is particularly novel. The method can be applied to other news serving systems. 
\item To facilitate live-data analysis of perceived news quality and of news quality scores, we develop quantitative and qualitative methods.
These methods can zoom in on individual news articles and their distribution, and also measure entire news ecosystems. Such analyses help both news publishers and consumers, which now depend on online news \cite{publishers2019fb}.
\end{itemize}

\begin{figure}[t]
\centering
\includegraphics[width=0.95\linewidth,trim={3cm 7cm 10cm 2cm},clip]{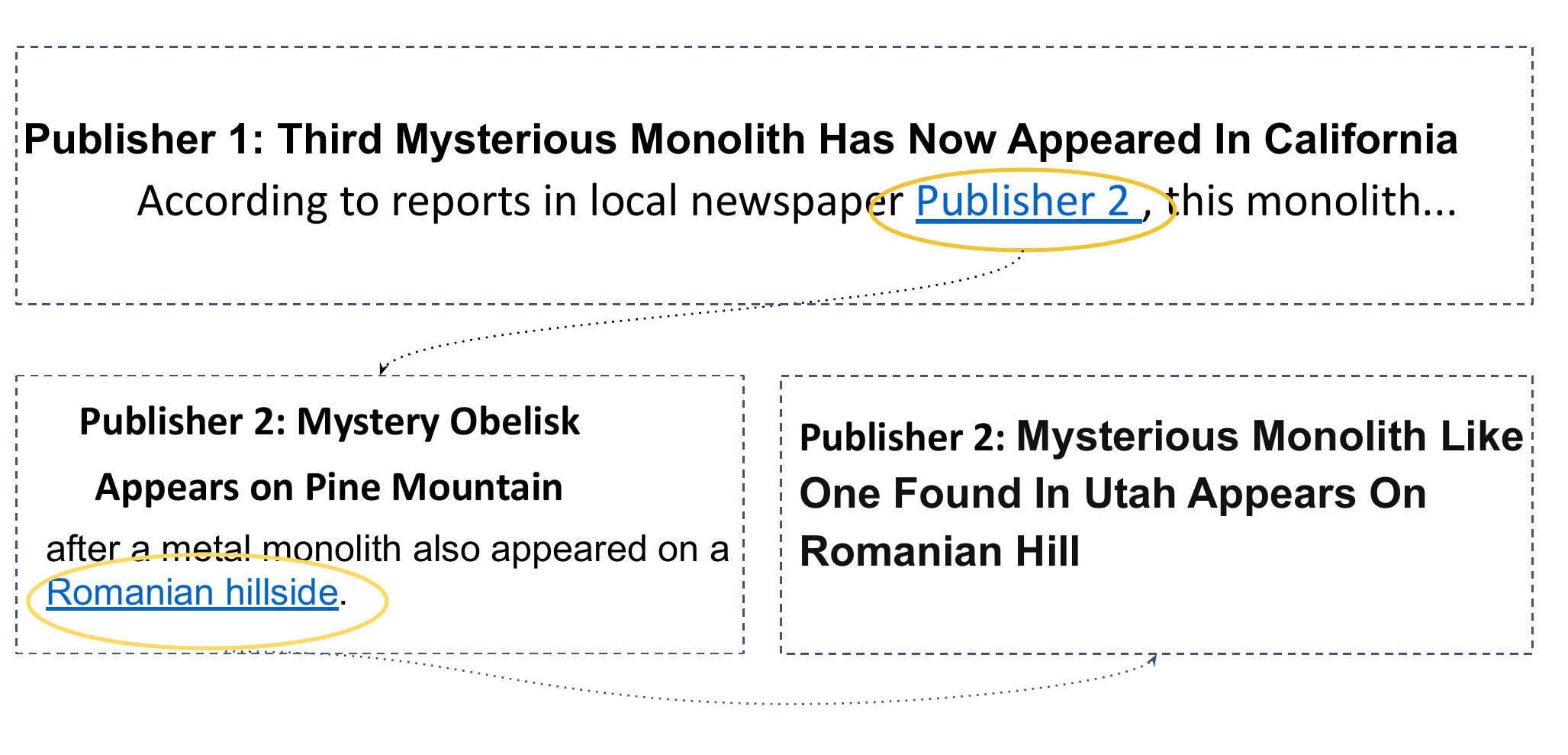}
\vspace{-4mm}
\caption{\label{fig:ex_citation}
Citations in news articles. The top snippet cites an article by another publisher. The cited article cites another article from the same publisher.}
\vspace{-3mm}
\end{figure}

\section{Background}
In this section, we first review the ideas behind PageRank and introduce the news citation graph. Then we outline ranking at Facebook, where we deploy our
originality signal. However, other social media use conceptually similar ranking systems and our contributions are not specific to Facebook.

\subsection{The News Citation Graph}
\label{sec:graph}

The PageRank algorithm was originally developed at Google to rank Web pages and sites to improve search results \cite{brin1998anatomy,cresci2015fame,cresci2017paradigm,page1999pagerank,ye2019mediarank}. Mathematically, it is a random-walk based algorithm to rank vertices in a graph. A Web page with many incoming links from large-weight web pages, has a greater weight. Page weights are propagated from each Web page to pages it links to.
In the news domain, the work by \citet{del2005ranking} introduced a related graph-based ranking algorithm where each vertex represents a news source, focusing on authoritative news sources and interesting news events.

\citet{ye2019mediarank} built an automated ranking system called MediaRank to rank news sources. They applied the PageRank algorithms on news reporting citation to rank news sources and proved that PageRank values are positively related to reporting quality measured by peer reputation and so on. \citet{zhang2018structured} introduced a set of signals for indicating the credibility of news collected from expert annotators. They grouped their indicators into two categories. The first group contains content indicators determined by the articles themselves --- mentions of organizations, studies, etc. Context indicators in the other group require analysis of external sources, such as author reputation
and/or recognition by peers in terms of the PageRank algorithm, as in
\citet{cresci2015fame,cresci2017paradigm}.

Similar to citations in academic papers, it is common to cite credible peers in the news industry, and such citations are important indicators of news source quality. Therefore, we introduce the news citation graph at news article level, instead of the domain level, to estimate the credibility of individual news articles. The idea is that when a news articleƒ is disproportionately cited by its peers, this indicates higher journalistic credibility. Whereas academic papers itemize their references and use reference numbers in citations, news articles follow a different style. In this work, we only consider citations in the form of links in a news article to other news articles.

Figure~\ref{fig:ex_citation} illustrates news article citations. The example at the top is from a news article by Publisher 1. This article cites multiple sources, one of them is shown: a news article from a another publisher, which cites another article by the same publisher. If a publisher breaks the story about an important news event, many other articles and publishers will cite it. 
 
We take snapshots of the news ecosystem  and index all our notation by time $t$ (Section \ref{sec:pagerank}). In particular,
$\mathcal{V}$ is the set of all news articles at time $t$, and $v\in \mathcal{V}$ denotes an individual article. We cluster such articles by news event or news story (Section \ref{sec:clustering}), denoting individual clusters $\mathcal{C} \subset \mathcal{V}$. When a news article $v$ cites another article $u$, we represent this by a directed edge $e_{v,u} \in \mathcal{E}$, where $\mathcal{E}$ is the set of edges in the citation graph. We also say that $e_{v,u}$ is $v$'s outbound edge and $u$'s inbound edge. Using these directed edges, we can compute the PageRank values of individual vertices (Section \ref{sec:pagerank}) by iteratively applying the following formula on every vertex in the graph in a topological order:
 
\begin{equation}
   n_v = \frac{1 - d}{|\mathcal{N}|} + d \sum_{u \in \mathcal{B}_v}\frac{n_u}{|\mathcal{B}_v|},
\end{equation}

where $n_v$ is the PageRank of article $v$ (initialized to 1) at time $t$,  $\mathcal{B}_v$ denotes the set of adjacent vertices (neighbors) of vertex $v$, $|\mathcal{B}_v|$ is the number of neighbors of $v$, and $d$ is a (constant) \textit{damping factor}, usually set to 0.85. The latter parameter dampens the propagation of weights through multiple edges.

\subsection{The News Article Representation}\label{sec:bert}

When estimating article originality, it is important to check how similar two articles are. Such checks are commonly
implemented with cosine similarity on vector embeddings. To produce necessary
embeddings, prior work uses the BERT (Bidirectional Encoder Representations from Transformers) network architecture \cite{devlin2018bert}, which achieved state-of-art results in many natural language processing tasks across different applications \cite{reddy2019coqa,lee2020biobert}. 
BERT handles previously unseen words by breaking them down into known subword fragments. It can also be updated on a regular basis to handle emerging keywords such as "COVID".
Original BERT models were DNNs pre-trained on the BooksCorpus \cite{zhu2015aligning} and the English Wikipedia. 
\hush{For example, a multilingual BERT implementation\footnote{\url{https://github.com/google-research/bert/blob/master/multilingual.md}} was trained on top 100 languages with Wikipedia data\footnote{\url{https://meta.wikimedia.org/wiki/List_of_Wikipedias}} to represent each news article based on its title. Using only titles conveniently neglects changes in article bodies, but emphasizes adequate handling of synonyms, rare words, and equivalent phrases --- BERT excels at these.} However, BERT networks can be specialized to a given use case by adding one dense layer and training it on adequate labeled data. Along these lines, \citet{reimers2019sentence} proposed a Sentence-BERT architecture that uses the Siamese network structure in the context of semantic similarity estimation.
\hush{on pair-wise labeled data (Section \ref{sec:clustering}).}
\hush{Whereas the pre-trained BERT embeddings are 768-dimensional, our dense layer reduces the dimensionality to 128.}

\subsection{News Feed Ranking} 
\begin{figure*}[ht]
\centering
\includegraphics[width=0.85\linewidth,trim={1cm 8cm 5cm 2.5cm},clip]{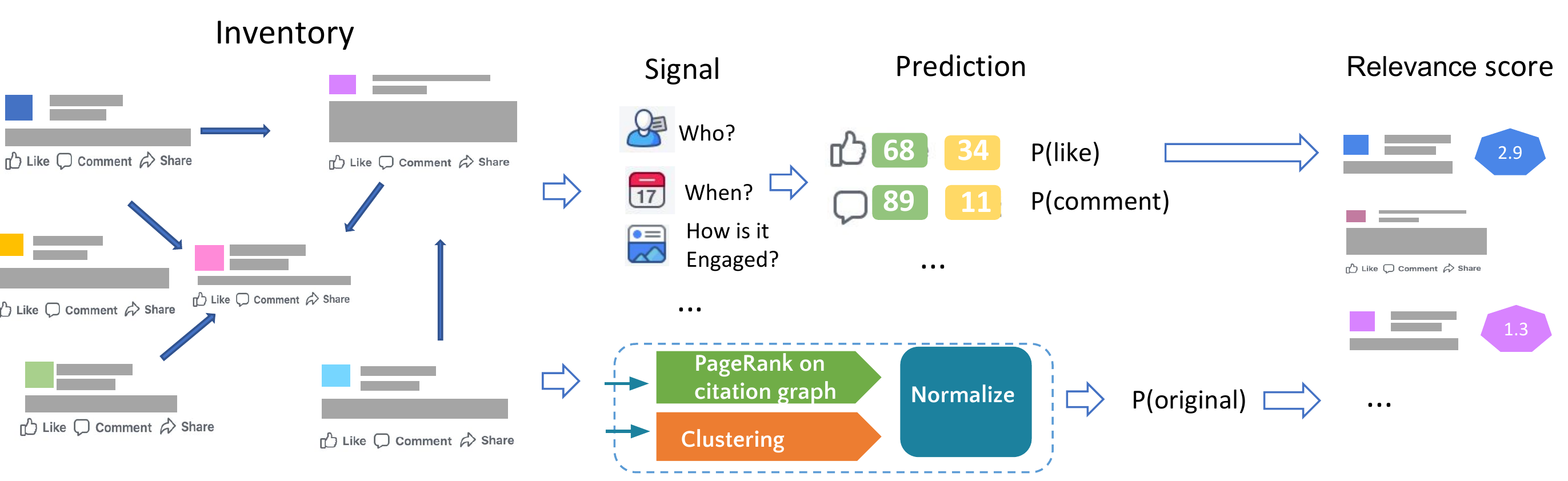}
\vspace{-3mm}
\caption{\label{fig:ranking} News Feed ranking at Facebook}
\vspace{-2mm}
\end{figure*}

\blue{The ranking of news has been extensively studied both in academia and industry \cite{del2005ranking,hu2006discovering,ye2019mediarank}. 
A number of publications in the information retrieval community
address this subject \cite{gwadera2009mining,kanhabua2011ranking,de2012chatter,tatar2014popularity,reis2015breaking,zheng2018drn,zhang2018structured}. In 2018, Nuzzle announced a ranking system for news sources called  NuzzleRank\footnote{\url{https://nuzzel.com/rank}} that
integrates various signals, including publisher authority information,
into a single score to rank news sources. }

\blue{Facebook's News Feed ranks not only news content, but also events from users' social graph~\cite{facebook,ranking2021fb}.
Ranking objectives optimize long-term user satisfaction, account for communities (friends and family, etc) \cite{people2018fb} and News Feed integrity  \cite{halevy2020integrity}(e.g., to discourage clickbait and prevent unlawful activities). When a user logs in to Facebook, they see their News Feed — which includes fresh updates from their friends, groups they joined, and pages they followed. News Feed ranking can be roughly divided into four stages: inventory, signals, prediction, and relevance scores \cite{facebook, ni2019feature, ranking2021fb}. 
Once a piece of content is posted, numerous signals are extracted — publication time, engagement counts, etc. Those signals are used to estimate the probabilities of possible individual user actions for each piece of content in the inventory, should they see it \cite{facebook,ni2019feature,ranking2021fb}. As a matter of notation, P(comment) represents the probability that a user comments on the update, while P(like) represents the probability that user  likes the content. At the last stage, we 
combine these predictions and
compute a ranking relevance score for each piece of content. Our news originality signal is deployed within this system summarized in Figure~\ref{fig:ranking}. News Feed ranking at Facebook incorporates many signals, and our originality signal enacts only subtle changes to the user experience as we explain later.}

\section{Problem Analysis}\label{sec:problem}
Here we examine the news originality landscape and motivate our work. Then we investigate the life-cycle of news stories on social media platforms. Understanding the news life-cycle is critical to deploying the originality signal within News Feed ranking.

\subsection{The Landscape of News Originality}

\begin{figure}[b]
\vspace{-5mm}
  \centering
  \includegraphics[width=0.9\columnwidth,trim={2cm 9cm 2cm 5cm},clip]{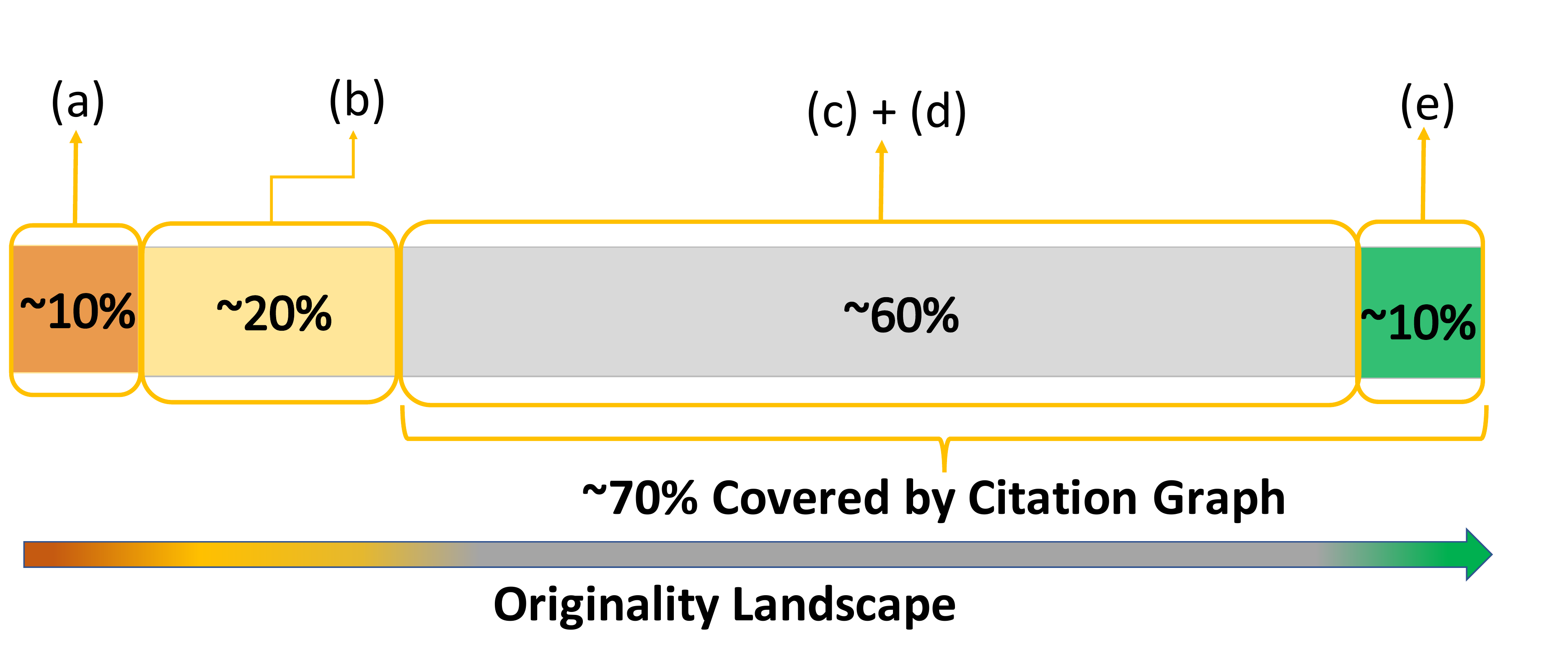}
  \vspace{-3mm}
\caption{\label{fig:landscape} News originality by bucket: (a) completely unoriginal; (b) highly unoriginal;
 (c) somewhat unoriginal; (d) potentially original but lacking peer recognition; (e) recognized as original by peers. For each bucket, we show estimated total views received by all news articles.}
  \vspace{-2mm}
\end{figure}

Our quantitative approach to news originality uses content buckets:

\begin{enumerate}[label=\alph*)]
\item \textit{completely unoriginal}, scraped or spun content with no editorial effort
\item \textit{highly unoriginal}, with very low editorial effort
\item \textit{somewhat unoriginal}, may be editorially produced but heavily cite other content without original reporting or analysis
\item \textit{potentially original but lacking peer recognition}
\item \textit{recognized as original by peers}: breaking news, eyewitness reports, exclusive scoop, investigative reporting, etc
\end{enumerate}

\begin{table}[b]
\vspace{-4mm}
    \centering
        \caption{\label{tab:ex_spun}
Examples of spun content. Publisher 1 posted original articles, while Publisher 2 replaced isolated words, phrases, and sentences in articles from Publisher 1.}
    \vspace{-2mm}
    \begin{tabular}{L{3.5cm} | L{3.5cm}}
 \toprule
         \sc Publisher 1 - Original & \sc Publisher 2 - Spun\\
          \midrule 
         Israel grants Rashida Tlaib West \underline{{\textbf{Bank}}} \underline{\textbf{visit}} on humanitarian grounds
  &  Israel grants Rashida Tlaib West \underline{\textbf{Financial Institution}} \underline{\textbf{go to}} humanitarian grounds  \\
    \hline
 Israel’s \underline{\textbf{interior}} minister on Friday said & Israel’s \underline{\textbf{inside}} minister on Friday said  \\
 \hline
 Pod \underline{\textbf{Foods gets}} VC backing to reinvent grocery distribution &  Pod \underline{\textbf{Meals will get}} VC backing to reinvent grocery distribution \\
  \bottomrule
    \end{tabular}
\end{table}

\textit{Scraped} content is copied from other sources without editorial efforts. \textit{Spun content} is taken from a post or a Web page, and posted with only minor modifications by humans or machines (see examples in Table~\ref{tab:ex_spun}). Common methods include paraphrasing, replacing words, and reordering paragraphs. By automating the \textit{spinning} of existing content one can quickly produce a large amount of content without scraping. 
\hush{Publisher 1 produces the original news article. Publisher 2 takes most of the content from the original one but replace the content with specific words, phrases, and sentences.} Scraped and spun content can eclipse original content and undermine its value, which warrants removal or limited distribution compared to original content.

\textit{Highly unoriginal} articles are produced by low-effort text changes. 
We find most of the news articles actually fall into the third bucket - 
\textit{somewhat unoriginal}. These articles may provide useful information,
but do not require much effort to produce.

\textit{Potentially original but lacking peer recognition} --- this bucket includes content that does not fit in earlier buckets and so may be original, but for various reasons does not receive peer recognition throughout the news cycle. Opinion pieces that receive little support often fall into this category. 
Thus, citation signals alone cannot distinguish between this bucket and unoriginal articles.

The \textit{highly original} news are  produced with significant effort to fact-check information and produce clear narratives, high-quality writing and visuals. Thoughtful and original news content is usually cited heavily by industry peers and contributes to
the reputation of individual content creators. Due to the effort and expertise required, the original news content are scarce.  Prioritizing the distribution of original content can help it reach greater audiences and benefits both the readers and the news industry in the long run~\cite{publishers2019fb}.

In general, it is difficult to judge each article for originality in isolation because this would require careful analysis of contents with the understanding of current events. Particularly challenging would be to distinguish rumors and fake news from reasonable reporting. Therefore, we draw additional insights from the news citation graph and the dynamics of online news. The special cases of scraped and spun content are handled by dedicated systems that are based on text hashing and fingerprinting, as well as text similarity metrics. In practice, such content does not appear in users' News Feed inventory and is therefore not treated in our work.%

\subsection{The Dynamics of Online News}\label{subsec:lifecycle}

News content published on the Internet can be easily indexed and archived, but 
it social media platforms tend to favor fresh news. That’s why news reporters strive to break a new story. To re-examine this conventional wisdom and determine how to reflect it in our work, we explore a large volume of news articles shared on Facebook and track the dynamics of user engagement metrics. We also visualize the life-cycle of typical online news stories and check the impact of adding valuable information days after the original publication.
\hush{Even if additional valuable information is added later, we usually do not observe another peak of engagement.}
As it turns out, the same pattern persists across different news categories --- world and local news, politics and entertainment news. 

\begin{figure}[b]
  \vspace{-2mm}
  \centering
  \includegraphics[width=0.95\linewidth,trim={4cm 4cm 9cm 1.5cm},clip]{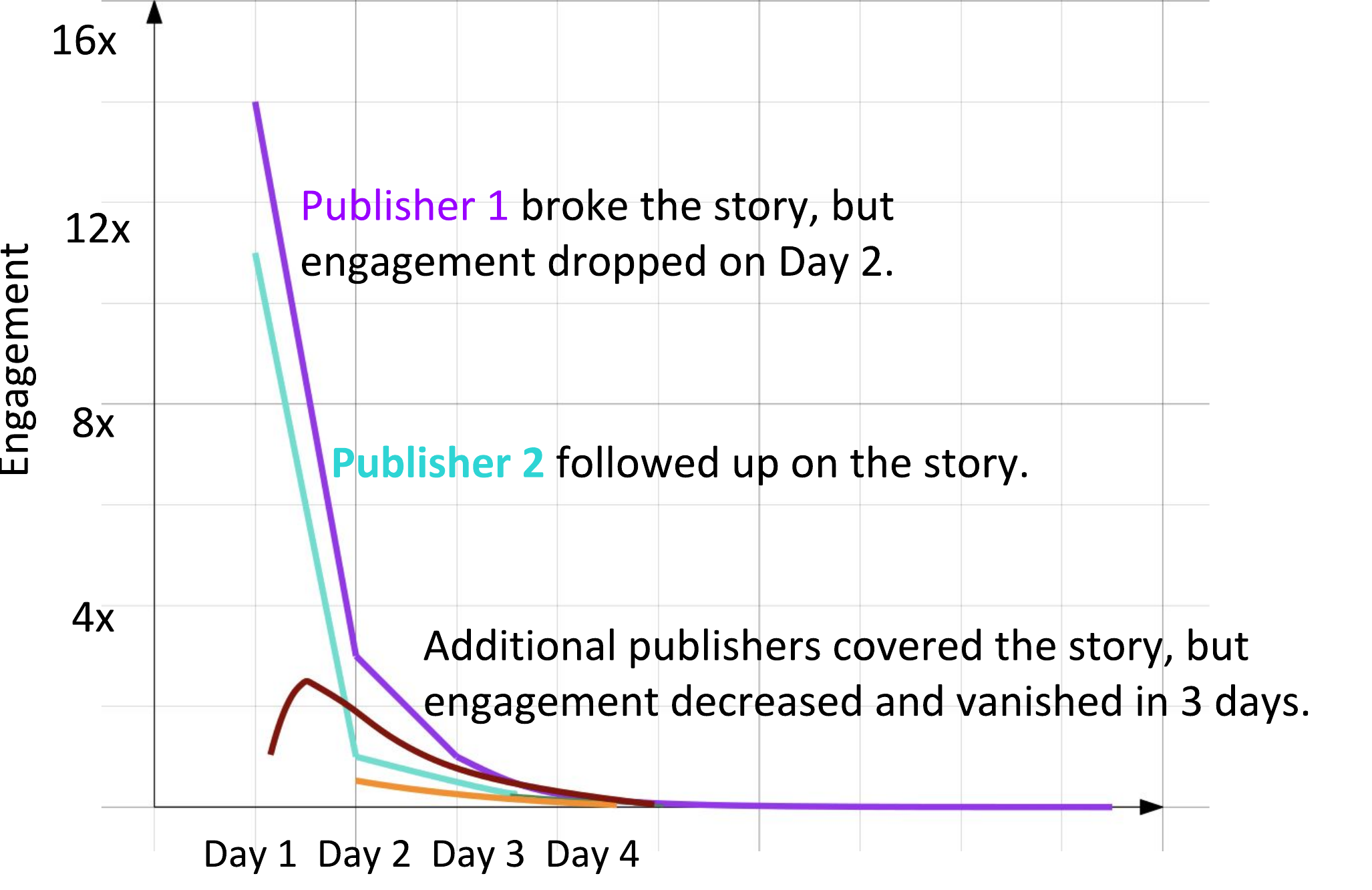}
  \vspace{-2mm}
  \caption{ \label{fig:lifecycle1} The life-cycle of a Spiderman story}
 \vspace{-3mm}
\end{figure}

Figure~\ref{fig:lifecycle1} illustrates how quickly users lose interest in a particular story. On September 27, 2019 Disney and Sony reached a deal for Spiderman movies, announcing that Spiderman would stay in the Marvel Universe. One publisher reported the story first. Almost 800 websites covered the news on the exact same day. On the second day, the engagement metrics of this story dropped significantly and eventually vanished on September 29, in just 3 days.

\begin{figure}[b]
\begin{tabular}{c}
\includegraphics[width=0.95\columnwidth,trim={1cm 0.5cm 8cm 0.8cm},clip]{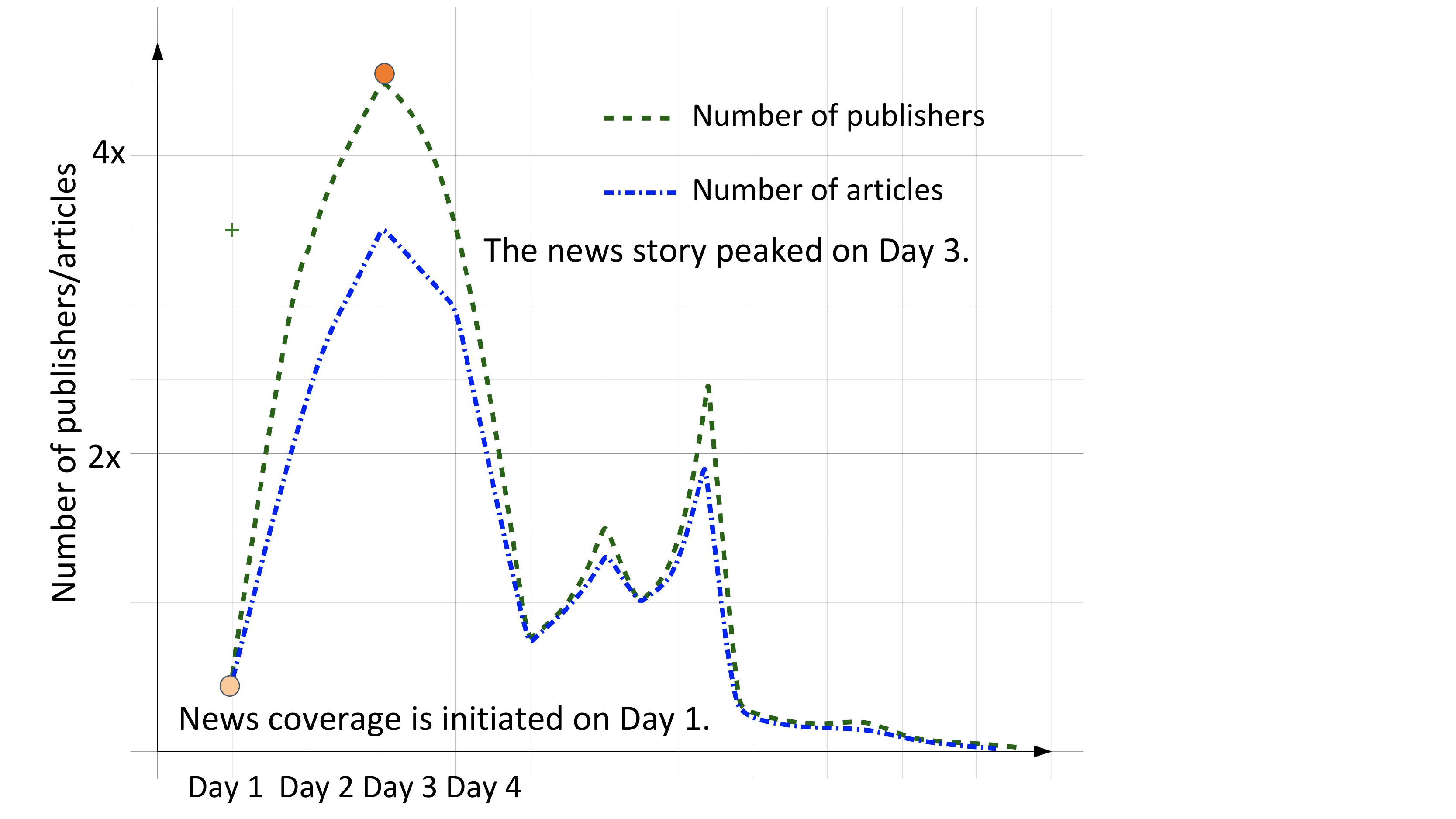}
      \vspace{-6pt}
      \\
      (a)
      \\
\includegraphics[width=0.95\columnwidth,trim={1cm 0.5cm 7cm 0.8cm},clip]{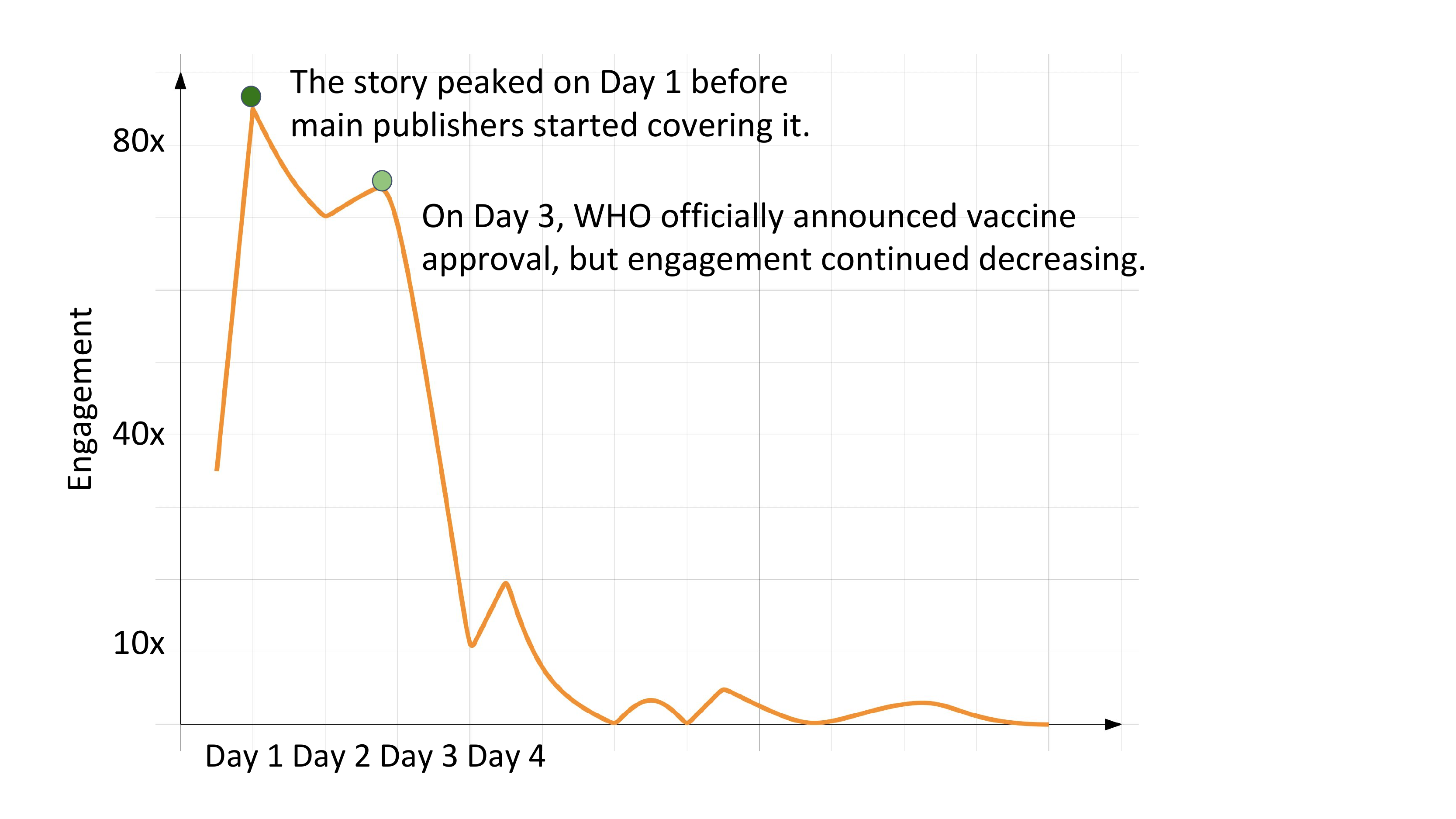} 
\vspace{-12pt}
\\
(b)
\end{tabular}
\vspace{-10pt}
\caption{\label{fig:lifecycle2} The life-cycle of the J\&J Ebola vaccine story}
\end{figure}

Figure~\ref{fig:lifecycle2} shows that adding information at a later time does not help gain traffic. On November 11, 2019 the Ebola vaccine by Johnson \& Johnson was approved. Our inventory showed that 17 websites published 34 related articles on that day, and user engagement metrics hit a peak. The news was first reported by a publisher who focuses on life science and medicine, which gained most traffic. Two days later, on November 13, the World Health Organization officially approved the vaccine. Many mainstream publishers covered this news, and we observed an inventory increase. However, this did not stimulate another engagement peak: traffic was mostly flat and almost vanished after seven days.  



Our data analysis suggests that ranking interventions can only be effective early in the life-cycle of a news story. 
This directly impacts the architecture and implementation of News Feed ranking,
posing challenges to both signal computation and ranking deployment. 
Therefore, we only focus on news articles published within the last seven days.
Our originality ranking intervention does not dramatically change users' News Feed experience because we do not alter the existing inventory of posts. However, the aggregated effect should reward publishers and reporters that produce thoughtful and original content. 

\section{Estimating News Originality}
\label{sec:idea}

Intuitively, \textit{news originality} refers to the process by which news content is created as well as the quality of news content. However, capturing these notions computationally appears challenging, especially when the content creation process remains opaque. Professional journalists and rates often find isolated text insufficient to rate originality and need additional context. Useful context includes ongoing news events and how much coverage they enjoyed, and also how a given news article is perceived by peers in the news ecosystem. A major precept in our work is that direct content analysis is neither sufficient nor necessary, whereas adequate context may provide sufficient signals to estimate originality.

To capture the context of individual news articles, we construct a \textit{news citation graph} (Section \ref{sec:graph}) for the entire news inventory at a fixed time. Peer recognition of each article is evaluated using the PageRank algorithm on this graph. An original piece of news could be cited by different publishers; it could also be a local news story cited by a major publisher with many subsequent citations --- both cases are captured adequately by PageRank. Here we emphasize the use of global PageRank values not restricted to particular news events. That is because quality articles often cite out-of-topic background material and may be cited under later news events.

We try to capture news ecosystem dynamics and emulate how professional raters or journalists estimate news originality level. To this end, PageRank values cannot be compared across topics and news events with very different amounts of news coverage. For a given news event or news story, we consider the entire news coverage as a cluster. Our insight is that {\em articles with the highest global PageRank values within each news-event cluster are most likely to be original}. Hence, we estimate news originality by normalizing global PageRank scores $n_v$ within each cluster $C_v$: \hush{(notation from Section \ref{sec:graph}).}

\vspace{-2mm}
\begin{equation}\label{eq:normalize}
s_v = \Big(\frac{n_v^p} {\sum_{u \in \mathcal{C}_v} n_u^p}\Big) ^ {\frac{1}{p}},\  
\hush{v \in \mathcal{C}_v, \ } p>0
\end{equation}
where $\mathcal{C}_v$ is the cluster of article $v$, and the $p$ constant defaults to $p=1.0$. Increasing $p$ would favor articles with higher $n_v$ values.

\begin{figure}[t]
  \centering
  \includegraphics[width=.93\columnwidth,trim={7cm 1.5cm 8cm 1cm},clip]{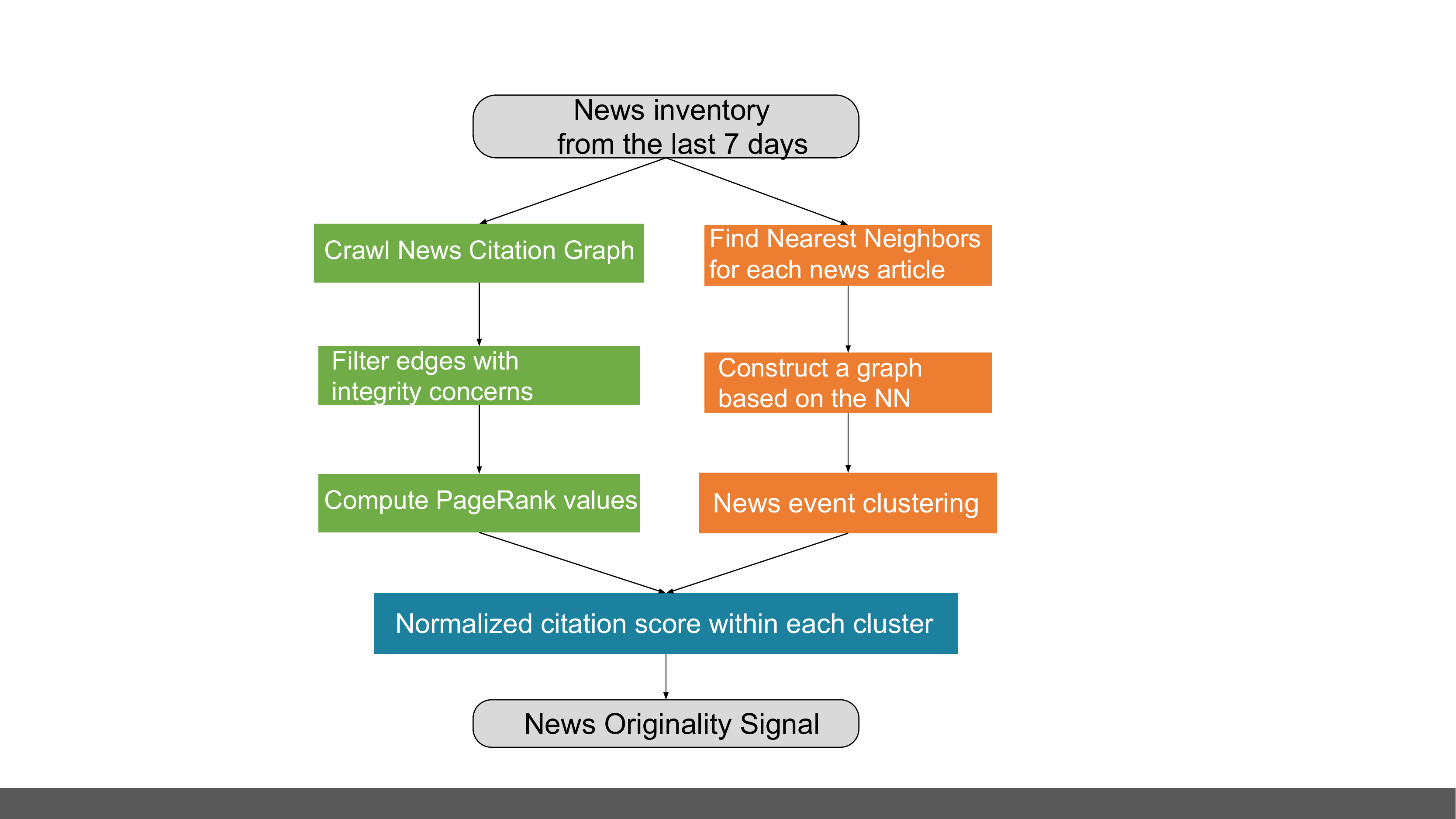}
  \vspace{-2mm}
  \caption{\label{fig:workflow} The workflow of our methodology.}
  \vspace{-4mm}
\end{figure}

Our process of estimating news originality is shown in Figure~\ref{fig:workflow}. Notably, we cannot evaluate a newly published article for originality before peers cite it. This introduces a delay and requires a near real-time system to deliver originality scores early in the news cycle. 

When using originality scores $s_v$ in News Feed ranking, we first convert them into P(original) $\in(0,1]$ as follows
\begin{equation}     \label{eq:linear}
  \mathrm{ P(original)} = \frac{\max(s_v, \theta) - \theta} { 1 - \theta}.
\end{equation}
Here $\theta\in (0,1]$ is the promotion threshold, i.e., only contents with $s_i^t > \theta$ can be promoted.
Then, we add P(original) to the relevance score as a second-order term: 
\begin{equation*}
   \mathrm{Relevance} = \ \alpha_1\  \cdot \mathrm{ P(comment)} \ +\ \alpha_2 \cdot \mathrm{P(share)}
     + \  \alpha_3\  \cdot \mathrm{P(like)}  
\end{equation*}
\vspace{-5mm}
\begin{equation}\label{eq:p_score}
      + \dots + \  \alpha_n \cdot \mathrm{ P(click)}\ \cdot \mathrm{ P(original)}. 
\end{equation}
Here P(comment), P(share), P(like) and P(click) are probabilities of respective events for the news article in question, and the weights $\alpha_i$ maximize long-term user satisfaction. Clearly, our originality signal is just one component of News Feed ranking that elevates peer-recognized content. Other signals elevate other content types.

\section{Implementation and Scaling}

Our preliminary investigation found that news articles highly cited by other articles tend to exhibit a higher level of originality. Therefore, we first build a citation graph of all news articles published in a seven-day window. Then, we calculate global PageRank values for individual articles, cluster news articles by news event/story in a scalable way, and normalize PageRank values within each cluster.

\subsection{PageRank with Integrity Considerations}
\label{sec:pagerank}
\blue{
We index all the news articles shared on the platform by leveraging the Facebook Crawler tool\footnote{\url{https://developers.facebook.com/docs/sharing/webmasters/crawler}}. The Facebook Crawler tool crawls the HTML of an app or website that was shared on Facebook via copying and pasting the link or by a Facebook social plugin. There are other open-source crawlers that serve the same purpose. Common Crawl\footnote{\url{https://commoncrawl.org/}} is a well-maintained open repository of Web crawl data that can be accessed and analyzed by anyone.}

We limit news articles in the graph to those posted within a seven-day moving window. After parsing the HTML, we traverse the output to get all {\tt <a>} tags, which define hyperlinks to other Web pages. Hyperlinks specified in the {\tt <a>} tag with different URLs may point to the same Web page. Therefore, we resolve alls URLs to canonical URLs\footnote{\href{https://developers.facebook.com/docs/sharing/webmasters/getting-started/versioned-link}{https://developers.facebook.com/docs/sharing/webmasters/getting-started/versioned-link}} and assign each news citation graph vertex a unique ID based on a canonical URL. If the cited Web page is also a recent news article, we establish an edge between the two news article vertices. With this news citation graph, we compute PageRank values for each news article. 


\blue{The raw citation graph is vulnerable to \textit{link farming}, as per \citet{du2007using}. That is, the graph may be manipulated by changing interconnected link structure of pages to add many inbound edges to a target page. To counter such manipulations, we disregard several types of citations before applying the PageRank algorithm.
As shown in  Figure~\ref{fig:ex_citation}, one typical example is \textit{self-linking} edges in $\mathcal{G}^t$ that cite an article published by the same publisher. Some Web sites link their articles to other Web sites without real content, but with automatic redirect to phishing sites or simply return to the citing article. 
These integrity filters mitigate the risk of manipulation. A filtered citation graph snapshot at each hour typically contains 300K--500K edges. The news articles without incoming and outgoing citations are excluded from the PageRank computation.
}
\red{
Despite their long history, attempts to manipulate PageRank in Web search have been successfully addressed \cite{penguin2016google}.
}

\blue{The original PageRank calculations work well with graphs that exhibit cycles, created when popular Web pages are revised to link to pages published later. Unlike the Web link graph, our news citation graph mostly contains links to past content since news posts on social networks are typically not revised. PageRank calculations simplify significantly on acyclic graphs and require a single linear-time graph traversal. However, in practice our citation graph contains enough cycles to question such simplifications.}

\subsection{News Event Clustering}\label{sec:clustering}

We now outline our clustering technique. As explained in Section \ref{sec:idea}, we normalize PageRank scores for individual news articles using PageRank scores of other articles in the same cluster. Intuitively, an important national news event and a local breaking news might carry similar amount of originality, but original articles in a larger cluster get more citations and higher PageRank scores. In addition to cluster normalization, computational scalability is also important --- on an uneventful day, our inventory snapshot contains 2M-3M articles, and we strive to process them in minutes.

\begin{figure}[ht]
  \centering
  \includegraphics[width=.65\linewidth,trim={6.5cm 0.5cm 15cm 0.1cm},clip]{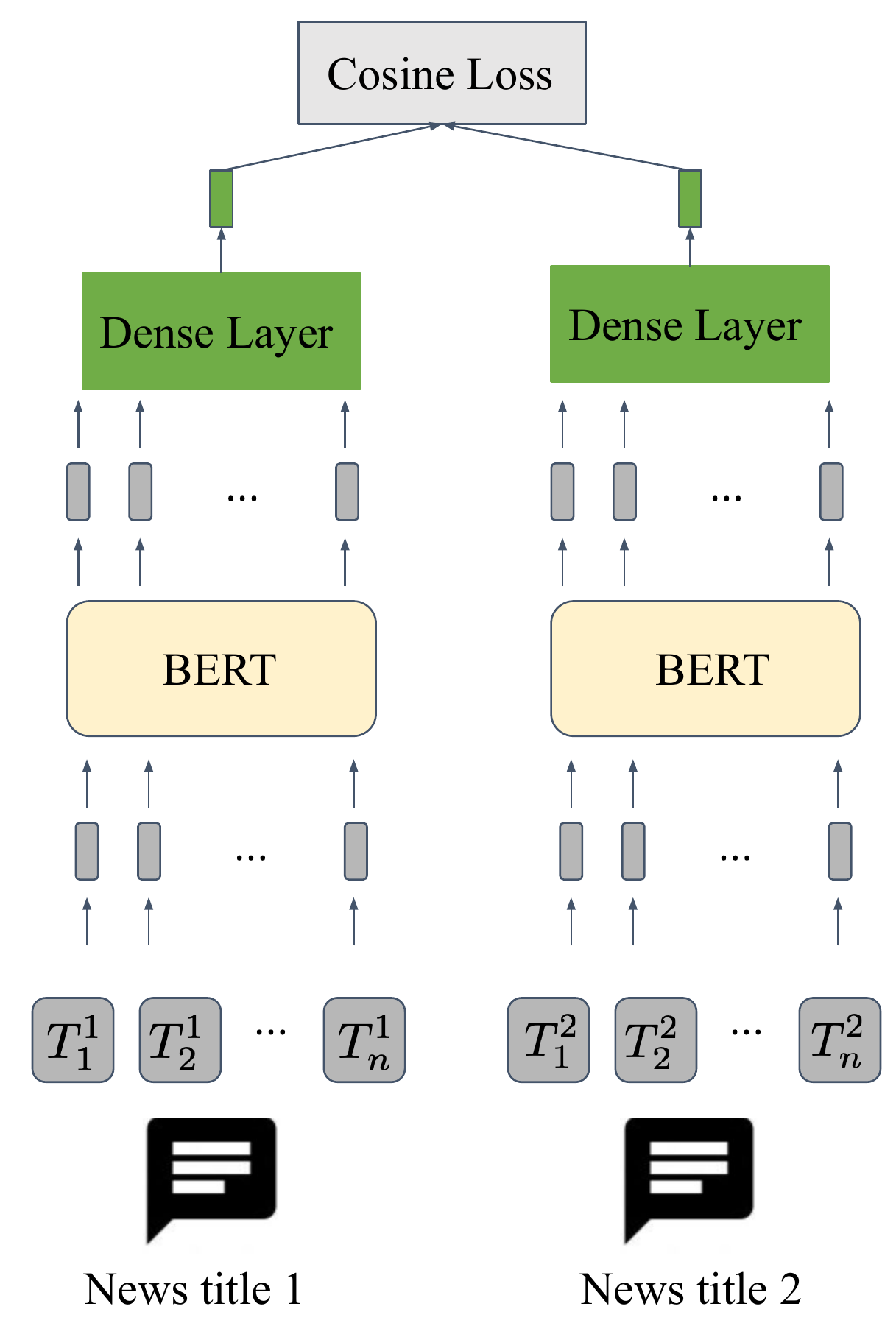}
  \vspace{-3mm}
  \caption{\label{fig:bert} Estimating sentence similarity using pre-trained BERT networks~\cite{reimers2019sentence}. The shared dense layer is trained.}
  \vspace{-5mm}
\end{figure}

We estimate the {\em topical similarity} of articles based on their titles, noting that articles with identical titles may have different PageRank scores.
We first lowercase article titles, remove punctuation and hash the titles to assemble duplicates into mini-clusters. For each unique title, we calculate a vector embedding based on the powerful and adaptable BERT DNN (Section~\ref{sec:bert}).
In addition to handling synonyms and equivalent phrases well, BERT also supports transfer learning.
To this end, we use a Siamese-twins network
architecture shown in Figure~\ref{fig:bert},
previously proposed for semantic similarity estimation~\cite{reimers2019sentence}. The two article titles are processed by the two constituent BERT models, which we implement in PyTorch using HuggingFace transformers \cite{wolf-etal-2020-transformers}. An
additional layer on top of BERT is a 128-dimensional fully connected (FC) layer with $tanh$ activation. 
In Figure~\ref{fig:bert}, $T_i$ 
represent the $i^{th}$ token in input sentences With the BERT network weights fixed, the top level is trained on labeled article pairs using the cosine embedding loss function $\mathcal{L}$
\begin{equation}
\mathcal{L}(x_1, x_2, y) = 
\begin{cases} 
1 - \cos(x_1, x_2) & \mbox{if } y = 1 \\ 
\max(0, \cos(x_1, x_2) - \mbox{ margin }) & \mbox{if } y = -1
\end{cases}
\end{equation}
where $x_1$ and $x_2$ represent the two input sentences respectively. $y = 1$ means the two sentences are same news event, while $y = -1$ means the two sentences are about completely different news event.  

\begin{figure}[b]
  \vspace{-3mm}
  \centering
  \includegraphics[width=.95\linewidth,trim={4.5cm 4cm 5cm 3.6cm},clip]{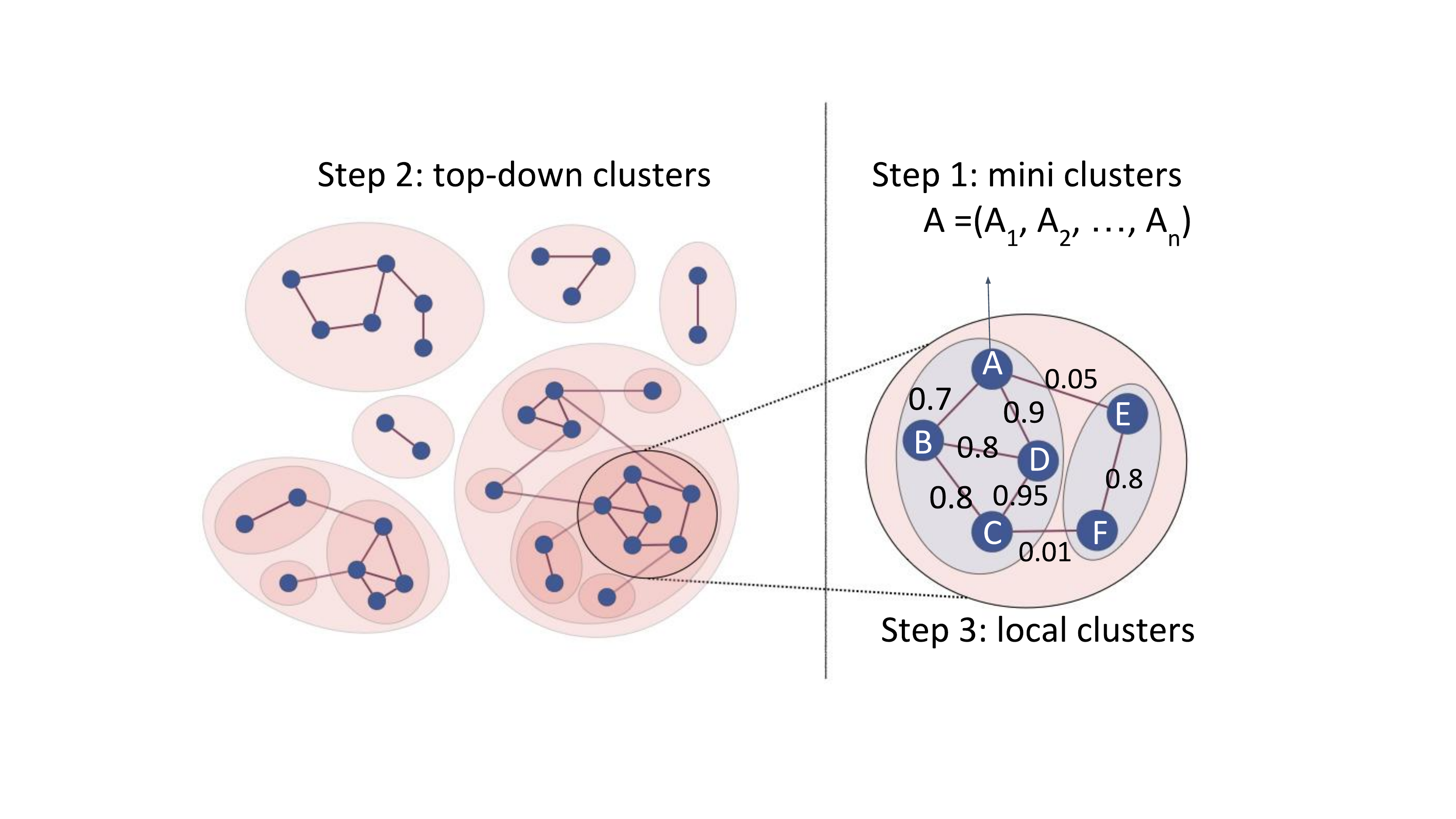}
  \vspace{-2mm}
  \caption{\label{fig:3step} Three-step clustering}
\end{figure}

BERT-based vector embeddings optimized to capture title similarity by cosine similarity support vector-based clustering algorithms. The choice of algorithms is driven by quality considerations and the ability to process millions of titles in several minutes, which we need to ensure frequent refresh of the news originality signal (in the context of  Section~\ref{subsec:lifecycle}).
\hush{
Every hour, over 2 millions news articles are viewed by users. As discussed in Section~\ref{subsec:lifecycle}, online news have pretty short life-cycle. Thus, our system needs to be as real-time as possible to capture news dynamics and use it in ranking effectively. We need to cluster all the articles fast and refresh the clusters frequently.
Thus, our algorithm should not only work well with vector representation (BERT embedding), but also be scalable and easy to improve and optimize. There are plenty of different clustering algorithms available. However, it's not easy to find one that satisfies all our needs.
}
 Clustering algorithms based on 
 K-Nearest-Neighbors (KNN) are a natural
 starting point, but specifying $K$ is not straightforward and for any given $K$ such algorithms risk producing inconsistent results in our application. Therefore, our three-step clustering in Figure~\ref{fig:3step} combines
 text hashing and KNN with greedy local search. Topical clusters often contain just a few different titles, while national news receive up to thousands citations per article. \hush{Global sparsity and tight local clusters are convenient for clustering algorithms.}

The set of unique article vectors is converted into an undirected KNN graph $\mathcal{G}$. For each vector, we find its $K=5000$ nearest neighbors based on {\em cosine similarity} (1 - {\em cosine distance}) and use cosine similarity for edge weights between adjacent vertices $v_i^t$ and $v_j^t$.
Lightweight edges are ignored, and subgraphs are defined by connected components of the resulting graph. Reasonable weight thresholds are found with a form of binary search guided by a subgraph size target. See details in Algorithm~\ref{algo:stage1}.

\begin{algorithm}
\SetAlgoLined
\KwInput{Weighted graph $\mathcal{G}=\{\mathcal{V}, \mathcal{E}\}$, subgraph target size $t$, optimization threshold $\epsilon$, $\ell$ = 0.0, $h = 1.0$}
\KwOutput{A set of subgraphs $\boldsymbol{\mathcal{S}}$ of approximately target size $t$}

  \DontPrintSemicolon
  \SetKwFunction{FMain}{findSubgraphs}
  \SetKwProg{Fn}{Function}{}{end}
  
  \Fn{\FMain{$\mathcal{G}$, $\epsilon$, $\ell$, $h$}:}{
  
  {$\boldsymbol{\mathcal{S}} = \emptyset$}
  
  \While{$h - \ell > \epsilon$}{
    $m = \frac{\ell + h}{2}$ \;
    {$\mathcal{G'} = \mathcal{G}$ without edges of weight < $m$ \;}
   {$\boldsymbol{\mathcal{C}}=$ 
   {\tt connectedComponents}($\mathcal{G}'$) \;}
    \ForEach{$c \in \mathcal{C}$}{
        \If{$|c| > b$}
        {
        Remove vertices in $c$ and their incident edges from $\mathcal{G}$

        $\boldsymbol{\mathcal{S}} = 
        \boldsymbol{\mathcal{S}}$ $\cup $
        \FMain{$c$, $\epsilon$, $l$, $m$}
        }
     }
     $h = m$
 }
{$\mathcal{G'} = \mathcal{G}$ without edges of weight < $m$ \;}
{$\boldsymbol{\mathcal{S}} = \boldsymbol{\mathcal{S}}$
  $\cup$
   {\tt connectedComponents}($\mathcal{G}'$) \;}
 \KwRet {$\boldsymbol{\mathcal{S}}$}\;
}

 \caption{ \label{algo:stage1}
  Split a graph into subgraphs with target size}
\end{algorithm}


  
 



An investigation of typical outputs of Algorithm~\ref{algo:stage1} suggested that clusters were generally reasonable, but local news and events with low coverage were not handled well. To remedy this deficiency, we form local clusters using greedy optimization to maximize the total edge weight $w_c$ inside clusters. We impart a default negative weight $\omega$ to pairs of vertices within a top-down cluster that are not connected by edges (not nearest neighbors). The smaller the $\omega$, the harder it is to create subclusters. For details, see Algorithm ~\ref{algo:stage2}.

\begin{example} Figure~\ref{fig:3step} illustrates local clusters in a subgraph: $\{A, B, C, D\}$ and $\{E, F\}$. Suppose $\omega=-0.1$. Then the total edge weight in cluster 1 is $w_1 = 0.7 + 0.8 + 0.8 + 0.9 + 0.95 - 0.1 = 4.15$ (no edge between $A$ and $C$), and in cluster 2 $w_2 = 0.8$. Although $A$ and $E$ are connected, the edge weight is so low that adding $E$ would not increase the total weight of cluster 1. The same reasoning applies to $F$. Therefore local clustering produces two clusters.
\end{example} 

 \begin{algorithm}
\SetAlgoLined
\KwInput{Weighted graph $\boldsymbol{g}$, negative weight $\omega$ for missing edges, number $R$ of independent randomized passes}
\KwOutput{An integer $c_v$ for each vertex $v$ (cluster assignment)}
 \RepTimes{$R$}{
 {Randomize the order of vertices in $\boldsymbol{g}$ } 
 Initialize each vertex $v$ in its own cluster $\boldsymbol{c}_v$ 

  \ForEach{$v \in \boldsymbol{g}$} {
        \ForEach{$u \in \mathcal{B}_v$} {
        {Try moving $v$ from cluster $c_v$ to cluster $c_u$} 
        {Add up internal weights for $c_u$ and $c_v$} 
        {Record $u$ with the highest sum of weights seen} 
        }
        {Move $v$ to maximize the sum of weights of $c_v$ and $c_u$}
    }
 \eIf{$\sum {w_c}$ increased}{
    repeat {\bf foreach} $v\in \boldsymbol{g} $}
 {Record the solution with the highest $\sum {w_c}$ seen}
}
 \caption{ \label{algo:stage2}
 Greedy local clustering}
\end{algorithm}

\subsection{Scalability}
 Building and processing the KNN graph with $K=5000$ nearest
 neighbors per vertex is a major performance bottleneck. On a typical day, all news articles from the last week fit in the RAM of a single server and can be processed reasonably quickly. However, this architecture is insufficiently scalable for the following reasons.
\begin{itemize}
    \item {\em Potential surges} of the news inventory during the election season, the New Year's Eve, etc.
    \item {\em Near real-time processing} benefits from additional compute resources (lower processing latency
    via using multiple servers).
    \item {\em Need for scaling to larger content inventory}.
    The challenge we are solving and our methods are fairly general, so can be applied to other social-network platforms that value originality. Now or in the future, such platforms may enjoy a much larger scale of content inventory.
\end{itemize}
 The overall design described in Section ~\ref{sec:clustering}
 naturally supports distributed processing to ensure greater overall scalability and robustness to surges. In fact, this is why
 Algorithm \ref{algo:stage1} performs {\em balanced} partitioning.
 Our implementation supports distributed clustering as well. We found that the upper bound on single-server capacity is an important parameter --- individual servers must receive a sufficient amount of work to justify distributed processing, but the data must fit into available RAM. Between the implied lower and upper bounds, there is a transition point where one can reduce the amount of computation at the cost of greater processing latency.

\section{Evaluation and Deployment}

Before deploying our news originality signal
to production at Facebook, we evaluate its functional components individually, evaluate the entire signal with the help of professional raters, then embed the signal into News Feed ranking and explore examples to check that everything works as expected. The production deployment is evaluated with an industry-standard technique ---an A/B test on live data for a limited subset of users before it is enabled for the main group of users~\cite{ab2015kdd}.

\subsection{Evaluation of Embeddings and Clustering}

In our rating flow, we ask professional raters to review pairs of news articles. The raters
assign a similarity level to each pair of articles: {\em different subjects}, {\em different subject but some common contents}, {\em same subjects with different aspects}, and {\em same subjects} (the four levels are explained in Table~\ref{tab:similarity_rating}).  For training, we collect 100K pairs of randomly sampled English news titles, using 40\% for finetuning, 10\%  for validation, and 50\% for test. Separately, we collect another 10K pairs of news articles to evaluate clustering performance. To sample likely-positive examples, we take some number of closest neighbors in terms of document embeddings and/or text similarity. Likely-negative samples are drawn from further-away neighbors that are sufficiently close to make the labeling task nontrivial.

\begin{table}[t]
    \centering
     \caption{\label{tab:similarity_rating}
    Guidelines for rating the similarity of article pairs}
     \vspace{-2mm}
    \begin{tabular}{L{1cm} | L{2.5cm} | L{3.5cm}}
 \toprule
    \sc Score & \sc Rating & \sc Criteria\\
          \midrule 
      0.0 & different subjects
  & the two articles cover completely different subjects \\
    \hline
1.0 & different subjects / some commonality & the two articles cover different subject but with share some content \\
 \hline
2.0 & same subject / different aspects & the two articles cover the same subject but
report different aspects of the same story \\
\hline
3.0 & same subject & the two articles cover the same subjects \\
  \bottomrule
    \end{tabular}
    \vspace{-1mm}
\end{table}


To compare our vector embeddings with FastText \cite{joulin2016fasttext} and Pytorch-BigGraph \cite{lerer2019pytorch} embeddings,
we represent similarity levels numerically by 0.0, 1.0, 2.0, 3.0 during training 
following Table~\ref{tab:similarity_rating}.
During evaluation, we binarize model scores
at thresholds 0.5, 1.5 and 2.5, then use
ROC AUC as the evaluation metric.
For example, AUC@0.5 considers article pairs with cosine similarity $\geq0.5$.
Table~\ref{tab:embedding} describes the
performance of our BERTPairwise model, which consistently outperforms pre-trained state-of-art embeddings. \hush{After the model is trained, we feed the titles of our candidate news articles to the model to generate the embedding of each news article.} 

\begin{table}[b]
\vspace{-3mm}
  \caption{
  \label{tab:embedding}
  The pairwise embedding vs. FastText \cite{joulin2016fasttext} and Pytorch-BigGraph \cite{lerer2019pytorch} embeddings}
  \vspace{-2mm}
\begin{tabular}{cccl}
    \toprule
    Model & AUC $@$ 0.5 (\%) &  AUC $@$ 1.5 & AUC $@$ 2.5\\
      \midrule
    FastText & 80.20 & 83.89 & 89.66 \\
    BigGraph & 82.95 & 84.87 & 89.61 \\
    \midrule
    BERTPairwise & 83.66 & 88.67 & 96.13\\
  \bottomrule
\end{tabular}
\end{table}


To evaluate our news-event clustering vs. human labels, we randomly sample 10K pairs of news articles in English from the candidate pool and send the pairs to professional annotators, along with guidelines in Table\ref{tab:similarity_rating}.  Then, we apply the clustering algorithms to the entire candidate pool. For each sampled pair, if the two articles appear the same cluster, the predicted label is positive, otherwise ---  negative. The clustering algorithm is evaluated
by precision and recall, then compared with
two well-known algorithms in Table~\ref{tab:clustering}.
{\tt DBSCAN} (density-based spatial clustering of applications with noise)\cite{ester1996density,schubert2017dbscan} is a highly scalable density-based algorithm. The {\tt Louvain} algorithm \cite{blondel2008fast} is one of the fastest and best-known community detection algorithms for large networks.

\begin{table}[t]
  \caption{  \label{tab:clustering}
 The performance of three-stage clustering with DBSCAN \cite{ester1996density} and the Louvain algorithm \cite{blondel2008fast}}
 \vspace{-2mm}
  \begin{tabular}{cccl}
    \toprule
\sc Algorithm & \sc Precision & \sc Recall \\
      \midrule
    DBSCAN & 43.07 & 73.04  \\
    Louvain & 81.01 & 47.57 \\
    Stage 1 + Louvain & 81.85 & 32.63\\ 

    \midrule
    three-stage clustering & \textbf{83.73} & \textbf{45.33}  \\
  \bottomrule
\end{tabular}
\vspace{-1mm}
\end{table}

\subsection{Evaluation by Professional Raters}

To assess the accuracy of our citations score signal, we sample the most viewed news articles identified as original, and the most viewed article not identified as original from the most viewed news domains over a seven-day period. Our professional raters have many years of news-industry experience and follow a deliberate process to ensure fair judgement for each article they rate on a three-point scale of news originality (Table~\ref{tab:original_rating}).
For the rating 3.0, our predicted labels match these results 90\% of the time. In other words, our signal attains 90\% accuracy in identifying original news.

\begin{table}[bt]
    \centering
        \caption{    \label{tab:original_rating}
Originality rating guidelines for human raters
        }
        \vspace{-2mm}
    \begin{tabular}{L{1cm} | L{2.5cm} | L{3.5cm}}
 \toprule
    \sc Score & \sc Rating & \sc Criteria \\
          \midrule 
      1.0 & unoriginal
  & borrows most of the content and language from other sources or is extremely thin / low information overall, and anything that is not properly syndicated. \\
    \hline
2.0 & possibly/somewhat unoriginal &  rewords borrowed content with its own language, but >70\% is borrowed OR properly syndicated \\
 \hline 
3.0 & fully original & 
is not a syndicated republishing,
little to no content is borrowed  \\
  \bottomrule
    \end{tabular}
 \vspace{-1mm}
\end{table}

\subsection{An Illustrative Example}
Besides the quantitative evaluation, we also performed qualitative case studies. Here we describe one example that illustrates how our system works.
On January 26, 2020, an article $n$ about the death of Kobe Bryant in a Calabasas helicopter crash was first reported by the publisher TMZ\footnote{TMZ: https://www.tmz.com}. In just 10 minutes, many publishers covered this story and cited TMZ. Over 200 articles fell into this news-event cluster, and the original story by TMZ ranked the highest. For such events, users would see news articles posted by the newspages they follow and shared by their friends. If the original news article is in a users' feed inventory, it gets prioritized. Note that our originality signal is only one component in the ranking formula. Users with preferences for certain publishers or strong affinity with their friends continue
seeing articles shared by those actors.

\subsection{Production Deployment and Evaluation}

\blue{The originality signal is intended for the relevance score calculation (see Figure~\ref{fig:ranking} and Equation~\ref{eq:p_score}) to increase the distribution of original news articles. 
To ensure its availability early in the news cycle, it is recalculated from scratch on an hourly basis. Building the news citation graph and news clusters takes only a few minutes, but system bottlenecks are observed in our current crawling infrastructure and in generating vector embeddings.
In practice, it takes time for the original articles to get cited, but running the workflow more often could find and promote original articles earlier. Such improvements are likely with further infrastructure optimization.}

\blue{Before making proposed changes to News Feed ranking at Facebook, we consulted with the academic and publishing communities and 
performed careful empirical evaluation.
In particular, we ran an A/B test on live data for several weeks, where the control group used prior production ranking rules and a small test group used revised ranking rules~\cite{ab2015kdd}. To estimate impact, we computed the increase in view counts at different thresholds  (Table~\ref{tab:stability}) and found 
our technique works well at different thresholds.}
 \blue{We have not observed statistically significant deteriorations in our proprietary metrics \cite{people2018fb,publishers2019fb,ranking2021fb} during the A/B test or after the subsequent full product launch. \hush{We have been tracking a goal metric called News Ecosystem Quality score. It is a synthetic score that combines several proprietary metrics such as clickbait prevalence. We observed a statistically significant score increase of \textbf{0.41\%} in our experiment.}  After additional checks and consultations, our signal was enabled for English-language content within Facebook's News Feed ranking system 
for most users in June 2020 ~\cite{newsorig2020fb}.
}

\red{Publishers may try to manipulate our news originality signal. To this end,
PageRank can be protected from abuse
\cite{penguin2016google}, wheras 
Facebook's integrity monitoring and enforcement \cite{halevy2020integrity} has a particular focus on coordinated inauthentic behaviors \cite{coord_inauth2021fb}. 
}

\hush{
To prevent unexpected regressions, data scientists at Facebook track hundreds of metrics before and after any changes to News Feed ranking. These metrics capture user engagement and user feedback, proxies for content quality and various aspects of News Feed integrity (such as clickbait prevalence).}
\begin{table}
  \caption{\label{tab:stability} User engagement lift in promoting original news}
  \vspace{-2mm}
  \begin{tabular}{cccl}
    \toprule
    \sc Originality threshold &
    \sc Increase in num. views (\%) \\
      \midrule
    0.4 & 15.36 \\
    0.5 & 14.72 \\
    0.6 & 14.30 \\
    0.7 & 13.83 \\
    0.8 & 13.38 \\
  \bottomrule
\end{tabular}
\vspace{-3mm}
\end{table}

\section{Conclusions and Perspectives}

In this paper, we introduce a strategy to prioritize original news in social networks. This strategy computes PageRank scores of news articles and estimates originality by normalizing PageRank scores for each news event. Equation \ref{eq:normalize} is a particularly novel contribution.

We deployed the originality signal to personalized Facebook News Feed, which compiles articles from sources followed by the user and user's friends \cite{facebook,news2021fb,ranking2021fb,publishers2019fb,people2018fb,ni2019feature}. When multiple articles are available in a user's inventory, we promote the more original ones. While subtle, such changes influence what the community sees. As part of our work, we performed conceptual, qualitative and quantitative evaluation to confirm that our techniques positively impact the news ecosystem. In particular, the exposure of original content has grown, and users received more content they liked. Over a longer timeframe, these developments should encourage publishers to invest more in original content.

\begin{acks}
We would like to thank Jon Levin,  Gabriella Schwarz, Lucas Adams, David Vickrey, Xiaohong Zeng, Joe Isaacson, Gedaliah Friedenberg,  Pengfei Wang, Feng Yan, Jerry Fu, Songbin Liu,  Yan Qi, Ranjan Subramanian, Adrian Le Pera, Vasu Vadlamudi, Julia Smekalina and others who supported and collaborated with us throughout.
\end{acks}

\bibliographystyle{ACM-Reference-Format}
\bibliography{main}


\begin{thebibliography}{37}


\ifx \showCODEN    \undefined \def \showCODEN     #1{\unskip}     \fi
\ifx \showDOI      \undefined \def \showDOI       #1{#1}\fi
\ifx \showISBNx    \undefined \def \showISBNx     #1{\unskip}     \fi
\ifx \showISBNxiii \undefined \def \showISBNxiii  #1{\unskip}     \fi
\ifx \showISSN     \undefined \def \showISSN      #1{\unskip}     \fi
\ifx \showLCCN     \undefined \def \showLCCN      #1{\unskip}     \fi
\ifx \shownote     \undefined \def \shownote      #1{#1}          \fi
\ifx \showarticletitle \undefined \def \showarticletitle #1{#1}   \fi
\ifx \showURL      \undefined \def \showURL       {\relax}        \fi
\providecommand\bibfield[2]{#2}
\providecommand\bibinfo[2]{#2}
\providecommand\natexlab[1]{#1}
\providecommand\showeprint[2][]{arXiv:#2}

\bibitem[\protect\citeauthoryear{Blondel, Guillaume, Lambiotte, and
  Lefebvre}{Blondel et~al\mbox{.}}{2008}]%
        {blondel2008fast}
\bibfield{author}{\bibinfo{person}{Vincent~D Blondel},
  \bibinfo{person}{Jean-Loup Guillaume}, \bibinfo{person}{Renaud Lambiotte},
  {and} \bibinfo{person}{Etienne Lefebvre}.} \bibinfo{year}{2008}\natexlab{}.
\newblock \showarticletitle{Fast unfolding of communities in large networks}.
\newblock \bibinfo{journal}{\emph{J. Statistical Mechanics: Theory and
  Experiment}} \bibinfo{volume}{2008}, \bibinfo{number}{10}
  (\bibinfo{year}{2008}), \bibinfo{pages}{10008}.
\newblock


\bibitem[\protect\citeauthoryear{Brin and Page}{Brin and Page}{1998}]%
        {brin1998anatomy}
\bibfield{author}{\bibinfo{person}{Sergey Brin} {and} \bibinfo{person}{Lawrence
  Page}.} \bibinfo{year}{1998}\natexlab{}.
\newblock \showarticletitle{The Anatomy of a Large-Scale Hypertextual Web
  Search Engine}.
\newblock \bibinfo{journal}{\emph{Computer Networks}}  \bibinfo{volume}{30}
  (\bibinfo{year}{1998}), \bibinfo{pages}{107--117}.
\newblock
\urldef\tempurl%
\url{http://www-db.stanford.edu/~backrub/google.html}
\showURL{%
\tempurl}


\bibitem[\protect\citeauthoryear{Brown and Levin}{Brown and Levin}{2020}]%
        {newsorig2020fb}
\bibfield{author}{\bibinfo{person}{Campbell Brown} {and} \bibinfo{person}{John
  Levin}.} \bibinfo{year}{2020}\natexlab{}.
\newblock \bibinfo{booktitle}{\emph{Prioritizing Original News Reporting on
  Facebook}}.
\newblock Facebook Newsroom.
\newblock
\urldef\tempurl%
\url{https://about.fb.com/news/2020/06/prioritizing-original-news-reporting-on-facebook/}
\showURL{%
\tempurl}


\bibitem[\protect\citeauthoryear{Cresci, Di~Pietro, Petrocchi, Spognardi, and
  Tesconi}{Cresci et~al\mbox{.}}{2015}]%
        {cresci2015fame}
\bibfield{author}{\bibinfo{person}{Stefano Cresci}, \bibinfo{person}{Roberto
  Di~Pietro}, \bibinfo{person}{Marinella Petrocchi}, \bibinfo{person}{Angelo
  Spognardi}, {and} \bibinfo{person}{Maurizio Tesconi}.}
  \bibinfo{year}{2015}\natexlab{}.
\newblock \showarticletitle{Fame for sale: Efficient detection of fake Twitter
  followers}.
\newblock \bibinfo{journal}{\emph{Decision Support Systems}}
  \bibinfo{volume}{80} (\bibinfo{year}{2015}), \bibinfo{pages}{56--71}.
\newblock


\bibitem[\protect\citeauthoryear{Cresci, Di~Pietro, Petrocchi, Spognardi, and
  Tesconi}{Cresci et~al\mbox{.}}{2017}]%
        {cresci2017paradigm}
\bibfield{author}{\bibinfo{person}{Stefano Cresci}, \bibinfo{person}{Roberto
  Di~Pietro}, \bibinfo{person}{Marinella Petrocchi}, \bibinfo{person}{Angelo
  Spognardi}, {and} \bibinfo{person}{Maurizio Tesconi}.}
  \bibinfo{year}{2017}\natexlab{}.
\newblock \showarticletitle{The paradigm-shift of social spambots: Evidence,
  theories, and tools for the arms race}. In \bibinfo{booktitle}{\emph{Proc.
  WWW}}. \bibinfo{publisher}{ACM}, \bibinfo{address}{Perth, Australia},
  \bibinfo{pages}{963--972}.
\newblock


\bibitem[\protect\citeauthoryear{De~Francisci~Morales, Gionis, and
  Lucchese}{De~Francisci~Morales et~al\mbox{.}}{2012}]%
        {de2012chatter}
\bibfield{author}{\bibinfo{person}{Gianmarco De~Francisci~Morales},
  \bibinfo{person}{Aristides Gionis}, {and} \bibinfo{person}{Claudio
  Lucchese}.} \bibinfo{year}{2012}\natexlab{}.
\newblock \showarticletitle{From chatter to headlines: harnessing the real-time
  web for personalized news recommendation}. In \bibinfo{booktitle}{\emph{Proc.
  WSDM}}. \bibinfo{publisher}{ACM}, \bibinfo{address}{Washington, USA},
  \bibinfo{pages}{153--162}.
\newblock


\bibitem[\protect\citeauthoryear{Del~Corso, Gulli, and Romani}{Del~Corso
  et~al\mbox{.}}{2005}]%
        {del2005ranking}
\bibfield{author}{\bibinfo{person}{Gianna~M Del~Corso},
  \bibinfo{person}{Antonio Gulli}, {and} \bibinfo{person}{Francesco Romani}.}
  \bibinfo{year}{2005}\natexlab{}.
\newblock \showarticletitle{Ranking a stream of news}. In
  \bibinfo{booktitle}{\emph{Proc. WWW}}. \bibinfo{publisher}{ACM},
  \bibinfo{address}{Chiba, Japan}, \bibinfo{pages}{97--106}.
\newblock


\bibitem[\protect\citeauthoryear{Devlin, Chang, Lee, and Toutanova}{Devlin
  et~al\mbox{.}}{2019}]%
        {devlin2018bert}
\bibfield{author}{\bibinfo{person}{Jacob Devlin}, \bibinfo{person}{Ming-Wei
  Chang}, \bibinfo{person}{Kenton Lee}, {and} \bibinfo{person}{Kristina
  Toutanova}.} \bibinfo{year}{2019}\natexlab{}.
\newblock \showarticletitle{{BERT}: Pre-training of Deep Bidirectional
  Transformers for Language Understanding}. In \bibinfo{booktitle}{\emph{In
  Proc. 17th NAACL}}. \bibinfo{publisher}{ACL}, \bibinfo{address}{Minneapolis,
  Minnesota}, \bibinfo{pages}{4171--4186}.
\newblock
\urldef\tempurl%
\url{https://doi.org/10.18653/v1/N19-1423}
\showDOI{\tempurl}


\bibitem[\protect\citeauthoryear{Du, Shi, and Zhao}{Du et~al\mbox{.}}{2007}]%
        {du2007using}
\bibfield{author}{\bibinfo{person}{Ye Du}, \bibinfo{person}{Yaoyun Shi}, {and}
  \bibinfo{person}{Xin Zhao}.} \bibinfo{year}{2007}\natexlab{}.
\newblock \showarticletitle{Using spam farm to boost PageRank}. In
  \bibinfo{booktitle}{\emph{Proc. Intl. Workshop on Adversarial Information
  Retrieval on the Web}}. \bibinfo{publisher}{ACM}, \bibinfo{address}{Banff
  Alberta, Canada}, \bibinfo{pages}{29--36}.
\newblock


\bibitem[\protect\citeauthoryear{Ester, Kriegel, Sander, Xu,
  et~al\mbox{.}}{Ester et~al\mbox{.}}{1996}]%
        {ester1996density}
\bibfield{author}{\bibinfo{person}{Martin Ester}, \bibinfo{person}{Hans-Peter
  Kriegel}, \bibinfo{person}{J{\"o}rg Sander}, \bibinfo{person}{Xiaowei Xu},
  {et~al\mbox{.}}} \bibinfo{year}{1996}\natexlab{}.
\newblock \showarticletitle{A density-based algorithm for discovering clusters
  in large spatial databases with noise.}. In \bibinfo{booktitle}{\emph{Proc.
  KDD}}, Vol.~\bibinfo{volume}{96}. \bibinfo{publisher}{AAAI Press},
  \bibinfo{address}{Portland, Oregon, USA}, \bibinfo{pages}{226--231}.
\newblock


\bibitem[\protect\citeauthoryear{Facebook}{Facebook}{2019}]%
        {publishers2019fb}
\bibfield{author}{\bibinfo{person}{Facebook}.} \bibinfo{year}{2019}\natexlab{}.
\newblock \bibinfo{booktitle}{\emph{People, Publishers, the Community}}.
\newblock Facebook Newsroom.
\newblock
\urldef\tempurl%
\url{https://about.fb.com/news/2019/04/people-publishers-the-community/}
\showURL{%
\tempurl}


\bibitem[\protect\citeauthoryear{Facebook}{Facebook}{2021a}]%
        {coord_inauth2021fb}
\bibfield{author}{\bibinfo{person}{Facebook}.}
  \bibinfo{year}{2021}\natexlab{a}.
\newblock \bibinfo{booktitle}{\emph{December 2020 Coordinated Inauthentic
  Behavior Report}}.
\newblock Facebook Newsroom.
\newblock
\urldef\tempurl%
\url{https://about.fb.com/news/2021/01/december-2020-coordinated-inauthentic-behavior-report/}
\showURL{%
\tempurl}


\bibitem[\protect\citeauthoryear{Facebook}{Facebook}{2021b}]%
        {news2021fb}
\bibfield{author}{\bibinfo{person}{Facebook}.}
  \bibinfo{year}{2021}\natexlab{b}.
\newblock \bibinfo{booktitle}{\emph{News content on Facebook}}.
\newblock Facebook Business Help Center.
\newblock
\urldef\tempurl%
\url{https://www.facebook.com/business/help/224099772719228}
\showURL{%
\tempurl}


\bibitem[\protect\citeauthoryear{Google}{Google}{2016}]%
        {penguin2016google}
\bibfield{author}{\bibinfo{person}{Google}.} \bibinfo{year}{2016}\natexlab{}.
\newblock \bibinfo{booktitle}{\emph{Penguin is now part of our core
  algorithm}}.
\newblock Google Search Central Blog.
\newblock
\urldef\tempurl%
\url{https://developers.google.com/search/blog/2016/09/penguin-is-now-part-of-our-core}
\showURL{%
\tempurl}


\bibitem[\protect\citeauthoryear{Gwadera and Crestani}{Gwadera and
  Crestani}{2009}]%
        {gwadera2009mining}
\bibfield{author}{\bibinfo{person}{Robert Gwadera} {and} \bibinfo{person}{Fabio
  Crestani}.} \bibinfo{year}{2009}\natexlab{}.
\newblock \showarticletitle{Mining and ranking streams of news stories using
  cross-stream sequential patterns}. In \bibinfo{booktitle}{\emph{Proc. ACM
  Conference on Information and Knowledge Management}}.
  \bibinfo{publisher}{ACM}, \bibinfo{address}{Hong Kong, China},
  \bibinfo{pages}{1709--1712}.
\newblock


\bibitem[\protect\citeauthoryear{Halevy et~al\mbox{.}}{Halevy
  et~al\mbox{.}}{2020}]%
        {halevy2020integrity}
\bibfield{author}{\bibinfo{person}{Alon Halevy} {et~al\mbox{.}}}
  \bibinfo{year}{2020}\natexlab{}.
\newblock \showarticletitle{Preserving Integrity in Online Social Networks}. In
  \bibinfo{booktitle}{\emph{Proc. KDD}}. \bibinfo{publisher}{ACM},
  \bibinfo{address}{USA}, \bibinfo{pages}{arXiv:2009.10311}.
\newblock


\bibitem[\protect\citeauthoryear{Hu, Li, Li, and Ma}{Hu et~al\mbox{.}}{2006}]%
        {hu2006discovering}
\bibfield{author}{\bibinfo{person}{Yang Hu}, \bibinfo{person}{Mingjing Li},
  \bibinfo{person}{Zhiwei Li}, {and} \bibinfo{person}{Wei-ying Ma}.}
  \bibinfo{year}{2006}\natexlab{}.
\newblock \showarticletitle{Discovering authoritative news sources and top news
  stories}. In \bibinfo{booktitle}{\emph{Asia Information Retrieval
  Symposium}}. \bibinfo{publisher}{Springer}, \bibinfo{address}{Beijing,
  China}, \bibinfo{pages}{230--243}.
\newblock


\bibitem[\protect\citeauthoryear{Joulin, Grave, Bojanowski, Douze, J{\'e}gou,
  and Mikolov}{Joulin et~al\mbox{.}}{2016}]%
        {joulin2016fasttext}
\bibfield{author}{\bibinfo{person}{Armand Joulin}, \bibinfo{person}{Edouard
  Grave}, \bibinfo{person}{Piotr Bojanowski}, \bibinfo{person}{Matthijs Douze},
  \bibinfo{person}{H{\'e}rve J{\'e}gou}, {and} \bibinfo{person}{Tomas
  Mikolov}.} \bibinfo{year}{2016}\natexlab{}.
\newblock \bibinfo{title}{FastText.zip: Compressing text classification
  models}.
\newblock
\newblock


\bibitem[\protect\citeauthoryear{Kanhabua, Blanco, and Matthews}{Kanhabua
  et~al\mbox{.}}{2011}]%
        {kanhabua2011ranking}
\bibfield{author}{\bibinfo{person}{Nattiya Kanhabua}, \bibinfo{person}{Roi
  Blanco}, {and} \bibinfo{person}{Michael Matthews}.}
  \bibinfo{year}{2011}\natexlab{}.
\newblock \showarticletitle{Ranking related news predictions}. In
  \bibinfo{booktitle}{\emph{Proc. Intl. ACM SIGIR Conference on Research and
  Development in Information Retrieval}}. \bibinfo{publisher}{ACM},
  \bibinfo{address}{Beijing, China}, \bibinfo{pages}{755--764}.
\newblock


\bibitem[\protect\citeauthoryear{Lada, Wang, and Yan}{Lada
  et~al\mbox{.}}{2021}]%
        {ranking2021fb}
\bibfield{author}{\bibinfo{person}{Akos Lada}, \bibinfo{person}{Meihong Wang},
  {and} \bibinfo{person}{Tak Yan}.} \bibinfo{year}{2021}\natexlab{}.
\newblock \bibinfo{booktitle}{\emph{How machine learning powers Facebook’s
  News Feed ranking algorithm}}.
\newblock Facebook.
\newblock
\urldef\tempurl%
\url{https://engineering.fb.com/2021/01/26/ml-applications/news-feed-ranking/}
\showURL{%
\tempurl}


\bibitem[\protect\citeauthoryear{Lee et~al\mbox{.}}{Lee et~al\mbox{.}}{2020}]%
        {lee2020biobert}
\bibfield{author}{\bibinfo{person}{Jinhyuk Lee} {et~al\mbox{.}}}
  \bibinfo{year}{2020}\natexlab{}.
\newblock \showarticletitle{BioBERT: a pre-trained biomedical language
  representation model for biomedical text mining}.
\newblock \bibinfo{journal}{\emph{Bioinformatics}} \bibinfo{volume}{36},
  \bibinfo{number}{4} (\bibinfo{year}{2020}), \bibinfo{pages}{1234--1240}.
\newblock


\bibitem[\protect\citeauthoryear{Lerer, Wu, Shen, Lacroix, Wehrstedt, Bose, and
  Peysakhovich}{Lerer et~al\mbox{.}}{2019}]%
        {lerer2019pytorch}
\bibfield{author}{\bibinfo{person}{Adam Lerer}, \bibinfo{person}{Ledell Wu},
  \bibinfo{person}{Jiajun Shen}, \bibinfo{person}{Timothee Lacroix},
  \bibinfo{person}{Luca Wehrstedt}, \bibinfo{person}{Abhijit Bose}, {and}
  \bibinfo{person}{Alex Peysakhovich}.} \bibinfo{year}{2019}\natexlab{}.
\newblock \showarticletitle{Pytorch-biggraph: A large-scale graph embedding
  system}.
\newblock \bibinfo{journal}{\emph{Proc. ML Sys Conference}}
  \bibinfo{volume}{1} (\bibinfo{year}{2019}), 12.
\newblock


\bibitem[\protect\citeauthoryear{Mosseri}{Mosseri}{2018a}]%
        {people2018fb}
\bibfield{author}{\bibinfo{person}{Adam Mosseri}.}
  \bibinfo{year}{2018}\natexlab{a}.
\newblock \bibinfo{booktitle}{\emph{Bringing People Closer Together}}.
\newblock Facebook Newsroom.
\newblock
\urldef\tempurl%
\url{https://about.fb.com/news/2018/01/news-feed-fyi-bringing-people-closer-together/}
\showURL{%
\tempurl}


\bibitem[\protect\citeauthoryear{Mosseri}{Mosseri}{2018b}]%
        {facebook}
\bibfield{author}{\bibinfo{person}{Adam Mosseri}.}
  \bibinfo{year}{2018}\natexlab{b}.
\newblock \bibinfo{booktitle}{\emph{News Feed Ranking in Three Minutes Flat}}.
\newblock Facebook Inc.
\newblock
\urldef\tempurl%
\url{https://newsroom.fb.com/news/2018/05/inside-feed-news-feed-ranking/}
\showURL{%
\tempurl}


\bibitem[\protect\citeauthoryear{Ni and other}{Ni and other}{2019}]%
        {ni2019feature}
\bibfield{author}{\bibinfo{person}{Xiuyan Ni} {and} \bibinfo{person}{other}.}
  \bibinfo{year}{2019}\natexlab{}.
\newblock \showarticletitle{Feature Selection for Facebook Feed Ranking System
  via a Group-Sparsity-Regularized Training Algorithm}. In
  \bibinfo{booktitle}{\emph{Proc. 28th ACM Intl. CIKM}}.
  \bibinfo{publisher}{ACM}, \bibinfo{address}{Beijing, China},
  \bibinfo{pages}{2085--2088}.
\newblock


\bibitem[\protect\citeauthoryear{Page, Brin, Motwani, and Winograd}{Page
  et~al\mbox{.}}{1999}]%
        {page1999pagerank}
\bibfield{author}{\bibinfo{person}{L. Page}, \bibinfo{person}{S. Brin},
  \bibinfo{person}{R. Motwani}, {and} \bibinfo{person}{T. Winograd}.}
  \bibinfo{year}{1999}\natexlab{}.
\newblock \bibinfo{booktitle}{\emph{The PageRank citation ranking: Bringing
  order to the Web.}}
\newblock \bibinfo{type}{{T}echnical {R}eport}. \bibinfo{institution}{Stanford
  InfoLab}.
\newblock


\bibitem[\protect\citeauthoryear{Reddy, Chen, and Manning}{Reddy
  et~al\mbox{.}}{2019}]%
        {reddy2019coqa}
\bibfield{author}{\bibinfo{person}{Siva Reddy}, \bibinfo{person}{Danqi Chen},
  {and} \bibinfo{person}{Christopher~D Manning}.}
  \bibinfo{year}{2019}\natexlab{}.
\newblock \showarticletitle{Coqa: A conversational question answering
  challenge}.
\newblock \bibinfo{journal}{\emph{Trans. ACL}}  \bibinfo{volume}{7}
  (\bibinfo{year}{2019}), \bibinfo{pages}{249--266}.
\newblock


\bibitem[\protect\citeauthoryear{Reimers and Gurevych}{Reimers and
  Gurevych}{2019}]%
        {reimers2019sentence}
\bibfield{author}{\bibinfo{person}{Nils Reimers} {and} \bibinfo{person}{Iryna
  Gurevych}.} \bibinfo{year}{2019}\natexlab{}.
\newblock \showarticletitle{Sentence-BERT: Sentence Embeddings using Siamese
  BERT-Networks}. In \bibinfo{booktitle}{\emph{Proc. of EMNLP}}.
  \bibinfo{publisher}{ACL}, \bibinfo{address}{Hong Kong, China},
  \bibinfo{pages}{3982--3992}.
\newblock
\urldef\tempurl%
\url{https://arxiv.org/abs/1908.10084}
\showURL{%
\tempurl}


\bibitem[\protect\citeauthoryear{Reis, Benevenuto, de~Melo, Prates, Kwak, and
  An}{Reis et~al\mbox{.}}{2015}]%
        {reis2015breaking}
\bibfield{author}{\bibinfo{person}{J. Reis}, \bibinfo{person}{Fabr{\i}cio
  Benevenuto}, \bibinfo{person}{Pedro~OS de Melo}, \bibinfo{person}{Raquel
  Prates}, \bibinfo{person}{Haewoon Kwak}, {and} \bibinfo{person}{Jisun An}.}
  \bibinfo{year}{2015}\natexlab{}.
\newblock \bibinfo{title}{Breaking the news: First impressions matter on online
  news}.
\newblock
\newblock


\bibitem[\protect\citeauthoryear{Schubert, Sander, Ester, Kriegel, and
  Xu}{Schubert et~al\mbox{.}}{2017}]%
        {schubert2017dbscan}
\bibfield{author}{\bibinfo{person}{Erich Schubert}, \bibinfo{person}{J{\"o}rg
  Sander}, \bibinfo{person}{Martin Ester}, \bibinfo{person}{Hans~Peter
  Kriegel}, {and} \bibinfo{person}{Xiaowei Xu}.}
  \bibinfo{year}{2017}\natexlab{}.
\newblock \showarticletitle{DBSCAN revisited, revisited: why and how you should
  (still) use DBSCAN}.
\newblock \bibinfo{journal}{\emph{ACM Trans. on Database Systems (TODS)}}
  \bibinfo{volume}{42}, \bibinfo{number}{3} (\bibinfo{year}{2017}),
  \bibinfo{pages}{1--21}.
\newblock


\bibitem[\protect\citeauthoryear{Tatar, Antoniadis, De~Amorim, and Fdida}{Tatar
  et~al\mbox{.}}{2014}]%
        {tatar2014popularity}
\bibfield{author}{\bibinfo{person}{Alexandru Tatar}, \bibinfo{person}{Panayotis
  Antoniadis}, \bibinfo{person}{Marcelo~Dias De~Amorim}, {and}
  \bibinfo{person}{Serge Fdida}.} \bibinfo{year}{2014}\natexlab{}.
\newblock \showarticletitle{From popularity prediction to ranking online news}.
\newblock \bibinfo{journal}{\emph{Social Network Analysis and Mining}}
  \bibinfo{volume}{4}, \bibinfo{number}{1} (\bibinfo{year}{2014}),
  \bibinfo{pages}{174}.
\newblock


\bibitem[\protect\citeauthoryear{Wolf et~al\mbox{.}}{Wolf
  et~al\mbox{.}}{2020}]%
        {wolf-etal-2020-transformers}
\bibfield{author}{\bibinfo{person}{Thomas Wolf} {et~al\mbox{.}}}
  \bibinfo{year}{2020}\natexlab{}.
\newblock \showarticletitle{Transformers: State-of-the-Art Natural Language
  Processing}. In \bibinfo{booktitle}{\emph{Proc. of EMNLP}}.
  \bibinfo{publisher}{ACL}, \bibinfo{address}{Online}, \bibinfo{pages}{38--45}.
\newblock
\urldef\tempurl%
\url{https://www.aclweb.org/anthology/2020.emnlp-demos.6}
\showURL{%
\tempurl}


\bibitem[\protect\citeauthoryear{Xu, Chen, Fernandez, Sinno, and Bhasin}{Xu
  et~al\mbox{.}}{2015}]%
        {ab2015kdd}
\bibfield{author}{\bibinfo{person}{Ya Xu}, \bibinfo{person}{Nanyu Chen},
  \bibinfo{person}{Addrian Fernandez}, \bibinfo{person}{Omar Sinno}, {and}
  \bibinfo{person}{Anmol Bhasin}.} \bibinfo{year}{2015}\natexlab{}.
\newblock \showarticletitle{From Infrastructure to Culture: A/B Testing
  Challenges in Large Scale Social Networks}. In
  \bibinfo{booktitle}{\emph{Proc. KDD}}. \bibinfo{publisher}{ACM},
  \bibinfo{address}{https://doi.org/10.1145/2783258.2788602},
  \bibinfo{pages}{2227–2236}.
\newblock


\bibitem[\protect\citeauthoryear{Ye and Skiena}{Ye and Skiena}{2019}]%
        {ye2019mediarank}
\bibfield{author}{\bibinfo{person}{Junting Ye} {and} \bibinfo{person}{Steven
  Skiena}.} \bibinfo{year}{2019}\natexlab{}.
\newblock \showarticletitle{MediaRank: Computational ranking of online news
  sources}. In \bibinfo{booktitle}{\emph{Proc. KDD}}. \bibinfo{publisher}{ACM},
  \bibinfo{address}{Anchorage, AK, USA}, \bibinfo{pages}{2469--2477}.
\newblock


\bibitem[\protect\citeauthoryear{Zhang et~al\mbox{.}}{Zhang
  et~al\mbox{.}}{2018}]%
        {zhang2018structured}
\bibfield{author}{\bibinfo{person}{A.~X. Zhang} {et~al\mbox{.}}}
  \bibinfo{year}{2018}\natexlab{}.
\newblock \showarticletitle{A structured response to misinformation: Defining
  and annotating credibility indicators in news articles}. In
  \bibinfo{booktitle}{\emph{Proc. WWW}}. \bibinfo{publisher}{ACM},
  \bibinfo{address}{Lyon, France}, \bibinfo{pages}{603--612}.
\newblock


\bibitem[\protect\citeauthoryear{Zheng et~al\mbox{.}}{Zheng
  et~al\mbox{.}}{2018}]%
        {zheng2018drn}
\bibfield{author}{\bibinfo{person}{Guanjie Zheng} {et~al\mbox{.}}}
  \bibinfo{year}{2018}\natexlab{}.
\newblock \showarticletitle{DRN: A deep reinforcement learning framework for
  news recommendation}. In \bibinfo{booktitle}{\emph{Proc. WWW}}.
  \bibinfo{publisher}{ACM}, \bibinfo{address}{Lyon, France},
  \bibinfo{pages}{167--176}.
\newblock


\bibitem[\protect\citeauthoryear{Zhu, Kiros, Zemel, Salakhutdinov, Urtasun,
  Torralba, and Fidler}{Zhu et~al\mbox{.}}{2015}]%
        {zhu2015aligning}
\bibfield{author}{\bibinfo{person}{Yukun Zhu}, \bibinfo{person}{Ryan Kiros},
  \bibinfo{person}{Rich Zemel}, \bibinfo{person}{Ruslan Salakhutdinov},
  \bibinfo{person}{Raquel Urtasun}, \bibinfo{person}{Antonio Torralba}, {and}
  \bibinfo{person}{Sanja Fidler}.} \bibinfo{year}{2015}\natexlab{}.
\newblock \showarticletitle{Aligning books and movies: Towards story-like
  visual explanations by watching movies and reading books}. In
  \bibinfo{booktitle}{\emph{Proc. IEEE Intl Conf. Computer Vision}}.
  \bibinfo{publisher}{IEEE Computer Society}, \bibinfo{address}{Santiago,
  Chile}, \bibinfo{pages}{19--27}.
\newblock


\end{thebibliography}


\end{document}